\documentclass[physics,article,submit,pdftex,moreauthors]{Definitions/mdpi} 
\firstpage{1} 
\pubvolume{1}
\issuenum{1}
\articlenumber{0}
\pubyear{2022}
\copyrightyear{2022}
\datereceived{} 
\dateaccepted{} 
\datepublished{} 
\hreflink{https://doi.org/} % If needed use \linebreak
\pdfoutput=1

\newcommand{\aap}{    {\it Astron. Astrophys.}}

\newcommand{\apj}{    {\it Astrophys. J.}}
\newcommand{\apjl}{   {\it Astrophys. J. Lett.}}

\newcommand{\mnras}{  {\it Mon. Not. Roy. Astron. Soc.}}

\newcommand{\pre}{    {\it Phys. Rev. E}}
\newcommand{\solphys}{{\it Solar Phys.}}
 
\newcommand{\ssr}{    {\it Space Sci. Rev.}} 
\newcommand{\araa}{    {\it Ann. Rev. Astron. Astrophys.}}

\newcommand{\pder}[2]{ \frac{\partial #1}{\partial #2} }
\newcommand{\pderN}[3]{ \frac{\partial^{#3} #1}{\partial #2^{#3}} }

\Title{Stability of slow magnetoacoustic and entropy waves in the solar coronal plasma with thermal misbalance}

\TitleCitation{Stability of slow magnetoacoustic and entropy waves in the solar coronal plasma with thermal misbalance}

\Author{Dmitrii Y. Kolotkov $^{1}$*, Valery M. Nakariakov $^{1,2}$ and Joseph B. Fihosy $^{1}$}

\AuthorNames{Dmitrii Y. Kolotkov, Valery M. Nakariakov and Joseph Fihosy}

\AuthorCitation{Kolotkov, D.; Nakariakov, V.; Fihosy, J.}
\address{%
$^{1}$ \quad  Centre for Fusion, Space and Astrophysics, Physics Department, University of Warwick, Coventry CV4 7AL, United Kingdom \\
$^{2}$ \quad  Centro de Investigacion en Astronomía, Universidad Bernardo O’Higgins, Avenida Viel 1497, Santiago, Chile}
\corres{Correspondence: D.Kolotkov.1@warwick.ac.uk}

\abstract{The back-reaction of the perturbed thermal equilibrium in the solar corona on compressive perturbations, also known as the effect of wave-induced thermal misbalance, is known to result in thermal instabilities chiefly responsible for the formation of fine thermal structuring of the corona. We study the role of the magnetic field and field-aligned thermal conduction in triggering instabilities of slow magnetoacoustic and entropy waves in quiescent and hot active region loops, caused by thermal misbalance. Effects of the magnetic field are accounted for by including it in the parametrisation of a guessed coronal heating function, and the finite plasma parameter $\beta$, in terms of the first-order thin flux tube approximation. Thermal conduction tends to stabilise both slow and entropy modes, broadening the interval of plausible coronal heating functions allowing for the existence of a thermodynamically stable corona. This effect is most pronounced for hot loops. In contrast to entropy waves, the stability of which is found to be insensitive to the possible dependence of the coronal heating function on the magnetic field, slow waves remain stable only for certain functional forms of this dependence, opening up perspectives for its seismological diagnostics in future.}


\begin{document}

%%%%%%%%%%%%%%%%%%%%%%%%%%%%%%%%%%%%%%%%%%
%\setcounter{section}{-1} %% Remove this when starting to work on the template.
%\section{How to Use this Template}

%The template details the sections that can be used in a manuscript. Note that the order and names of article sections may differ from the requirements of the journal (e.g., the positioning of the Materials and Methods section). Please check the instructions on the authors' page of the journal to verify the correct order and names. For any questions, please contact the editorial office of the journal or support@mdpi.com. For LaTeX-related questions please contact latex@mdpi.com.%\endnote{This is an endnote.} % To use endnotes, please un-comment \printendnotes below (before References). Only journal Laws uses \footnote.

% The order of the section titles is different for some journals. Please refer to the "Instructions for Authors” on the journal homepage.

\section{Introduction}
\label{sec:intro}

Magnetohydrodynamic (MHD) waves which appear as perturbations of various macroscopic plasma parameters, is one of the most intensively studied physical phenomena in the corona of the Sun \citep[e.g.,][]{2020ARA&A..58..441N}. Coronal waves are considered as possible agents responsible for heating of the coronal plasma \citep[e.g.][]{2020SSRv..216..140V}, and also as natural probes of the medium in the method of MHD seismology \citep[e.g.][]{2012RSPTA.370.3193D}.   
The waves detected in the corona in various observational bands can be clearly distinguished by the propagation speed. Coronal slow magnetoacoustic waves have the propagation speed about the local sound speed which is determined by the plasma temperature, i.e., from several tens to a few hundred km\,s$^{-1}$.  The slow waves appear mainly in two, propagating and standing regimes. Propagating slow waves are usually observed as EUV intensity disturbances running upwards along the apparent direction of the magnetic field in magnetically open plasma structures \citep[e.g.,][]{2021SSRv..217...76B}, e.g., polar plumes, as well as in legs of coronal loops and plasma fans \citep[e.g.,][]{2009SSRv..149...65D}. Typical oscillation periods are from a few to several minutes. Propagating slow waves are detected up to a certain height, and the downward propagating waves have not been detected. 

Slow waves in the other, standing regime, also called \lq\lq SUMER\rq\rq\ oscillations, are usually detected as periodic Doppler shifts of coronal emission lines and sometimes the periodic variation of the line intensity \citep[see][for a recent comprehensive review]{2021SSRv..217...34W}.
Oscillation periods of standing slow waves are typically longer than in the propagating regime \citep[e.g.][]{2019ApJ...874L...1N}, reaching thirty minutes \citep[e.g.][]{2017ApJ...847L...5P}. This difference could be readily attributed to different mechanisms for the periodicity. If for propagating waves the oscillation period is prescribed by the driver, the oscillation period of standing waves in determined by the resonant period of the oscillating plasma structure. For slow waves, this period is determined by the ratio of the loop's length to the sound speed. The latter is proportional to the root of the plasma temperature.
The damping time is about the oscillation period. The damping time scales with the oscillation period almost linearly \citep{2011SSRv..158..397W}. Moreover, the oscillation quality factor defined as the ratio of the damping time to the oscillation period is found to decrease with the increase in the relative amplitude, as the amplitude to the minus two thirds \citep{2019ApJ...874L...1N}. 
The nature of the possible third, sloshing or reflecting form of coronal slow waves is still intensively debated \citep[e.g.][]{2013ApJ...779L...7K, 2016ApJ...826L..20R, 2021ApJ...914...81K}.  

Since works of \citet{1999ApJ...514..441O} and \citep{2002ApJ...580L..85O}, theoretical modelling of coronal slow waves has usually been performed in term of ideal or dissipative MHD \citep[e.g.][]{2000A&A...362.1151N, 2001A&A...379.1106T, 2003A&A...408..755D, 2005A&A...436..701S, 2013A&A...553A..23R, 2015ApJ...807...98Y}. However, this approach cannot be considered as complete, as it is necessary to take into account not only perturbations of the mechanical equilibrium, but also of the thermal equilibrium. The coronal thermal equilibrium is fulfilled by the balance of energy losses by radiation and thermal conduction, and energy gains by an unknown yet coronal heating mechanism. A compressive perturbation such as a slow wave, causes variations of macroscopic parameters of the plasma, which, in turn, modify the efficiency of the energy losses and gains mechanisms. As the radiative losses, thermal conduction and heating are not likely to depend on the plasma density, temperature and the magnetic field strength identically, the wave causes thermal misbalance in the corona. The back-reaction of the misbalance on the compressive perturbation, resulting in thermal instability, has been subject to intensive studies for several decades \citep[e.g.,][]{1965ApJ...142..531F, 1991SoPh..134..247V, 1992PPCF...34..411H}. In the solar coronal context, this instability is believed to lead to prominence formation, coronal rain, and long-period thermal cycles \citep[e.g.,][]{2022FrASS...920116A}. However, the stable regime is also of interest, as it allows us to interpret slow wave motions observed in the corona, and use them for coronal plasma diagnostics. In some studies of stable coronal waves,the radiative losses and background heating have been taken into account too. However, the heating function was usually taken to be independent of the plasma parameters, aiming to maintain the thermal equilibrium \citep[e.g.][]{2004A&A...422..351T}. 

The effect of the back-reaction of the wave-induced thermal misbalance on coronal MHD waves has recently become subject to intensive studies. In particular, the misbalance was found to cause coupling of slow magnetoacoustic and entropy waves \citep{2021SoPh..296...96Z}. Occurrence of two characteristic time scales of the misbalance leads to the previously unaccounted wave dispersion \citep{2019PhPl...26h2113Z}. The dispersion results in the dependence of the effective adiabatic index, and the wave phase and group speeds on the wave frequency, and a formation of an oscillatory pattern in an initially broadband slow wave. The characteristic oscillation period is determined by the derivatives of the combined radiation loss and heating function with respect to the density and temperature, evaluated at the equilibrium. Interestingly, in coronal loops, the values of these two characteristic time scales of the misbalance are of the same order of magnitude as the slow wave oscillation periods, detected in observations. In particular, it suggests that the effect could be responsible for long-period quasi-periodic pulsations detected in the decay phases of solar and stellar flares \citep{2021SSRv..217...66Z}. Recently, \citet{2022MNRAS.514L..51K} used the thermal misbalance effect for explaining the observed scaling of the standing slow oscillation damping time with the oscillation period. In addition, this effect modifies phase relations between various plasma parameters perturbed by coronal slow waves \citep{2021SoPh..296..105P, 2022SoPh..297....5P}. The effect of thermal misbalance can also play an important role in nonlinear dynamics of coronal slow waves \citep{2020PhRvE.101d3204Z}, and in fragmentation of coronal current sheets \citep{2021SoPh..296...74L}. Another important effect of the thermal misbalance is the modification of the slow wave damping pattern. The wave can experience additional attenuation, or, contrarily, amplification \citep{2017ApJ...849...62N, 2019A&A...628A.133K}. In the latter case, the coronal plasma behaves as an active medium \citep{2021PPCF...63l4008K} c.f. the gain medium in a laser. The energy which compensates the radiative and conductive losses is supplied by the coronal heating mechanism, i.e., the corona is a thermodynamically open system. This property provides us with a tool for seismological probing of the coronal heating mechanism \citep{2020A&A...644A..33K}.  

The aim of this paper is to generalise the study of \citep{2020A&A...644A..33K} to the regime with finite thermal conduction along the field and a finite plasma $\beta$. The paper is organised as follows. In Section~\ref{sec:model}, we present the main assumptions, governing equations, and the dispersion relation. Section~\ref{sec:thermal_cond} addresses the stabilisation of thermal instability by finite thermal conduction. In Section~\ref{sec:beta}, we discuss the effect of finite plasma $\beta$, i.e., go beyond the infinite field approximation. The final section summarises the results obtained, and presents discussion of their implications. 

\section{Governing equations and dispersion relation}
\label{sec:model}

Neglecting the effects of waveguide dispersion, slow magnetoacoustic waves can be described in terms of the first-order thin flux tube approximation with finite plasma-$\beta$ \citep{1996PhPl....3...10Z}. We neglect the magnetic field twist, assuming that the field is parallel to the axis of the magnetic flux tube, directed in the $z$ direction. In the linear regime and accounting for thermal conduction and thermal misbalance, the governing equations take the form \citep{2017ApJ...849...62N, 2021A&A...646A.155D, 2021PPCF...63l4008K, 2021SoPh..296..122B},
\begin{align} \label{eq:continuity}
	&\pder{\rho_1}{t} +2 \rho_0 v_{r1} + \rho_0 \pder{u_1}{z}  =0,\\
	&\rho_0 \pder{u_1}{t} + \pder{P_1}{z}  =0,\\
	&C_\mathrm{V} \rho_0 \pder{T_1}{t} - \frac{k_{B} T_0}{m}  \pder{\rho_1}{t} = - \rho_0 \left(Q_{\rho}\rho_1 + Q_{T} T_1 + Q_{B} B_\mathrm{1} \right)  + \kappa_\parallel \pderN{T_1}{z}{2},\label{eq:energy}\\
	&P_1  - \frac{k_{B} }{m} \left(\rho_0 T_1 +T_0 \rho_1 \right)  =0,\\
	&P_1+\frac{B_0 B_1}{\mu_0}  = 0,\\
	&\pder{B_1}{t} + 2 B_0 v_{r1}   =0.\label{eq:induction}
\end{align}
In Eqs.~(\ref{eq:continuity})--(\ref{eq:induction}), $\rho_1$, $u_1$, $P_1$, $T_1$, and $B_1$ stand for the small-amplitude perturbations of the loop's plasma density, parallel velocity (along the $z$ axis), gas pressure, temperature, and magnetic field, respectively; $v_{r1}$ is the radial {(in cylindrical geometry)} derivative of the radial plasma velocity taken at the loop's axis which accounts for a weak obliquity of slow waves; the subscripts \lq\lq 0\rq\rq\ indicate the unperturbed values of those quantities; and $\gamma$, $m$, $k_\mathrm{B}$, and $C_\mathrm{V} = (\gamma-1)^{-1}k_\mathrm{B}/m$ stand for the standard adiabatic index (5/3), mean particle mass (0.6 of the proton mass), Boltzmann constant, and specific heat capacity, respectively.
As such, characteristic speeds in the model described by Eqs.~(\ref{eq:continuity})--(\ref{eq:induction}) are the sound speed, $c_\mathrm{s}^2={\gamma k_\mathrm{B}T_0/m}$ and the tube speed, $c_\mathrm{T}^2={c_\mathrm{s}^2c_\mathrm{A}^2/(c_\mathrm{s}^2 + c_\mathrm{A}^2)}$, with $c_\mathrm{A}^2=B_0^2/\mu_0\rho_0$ being the Alfv\'en speed (for a low-$\beta$ plasma, $c_\mathrm{T} \to c_\mathrm{s}$). There are no steady flows in the equilibrium state.

{Weak obliquity with respect to the guiding magnetic field is an intrinsic property of slow magnetoacoustic waves in a magnetic cylinder in the regime of finite plasma-$\beta$ \citep[e.g.][]{2006RSPTA.364..447R}. Indeed, in the infinite magnetic field approximation in which plasma-$\beta$ is assumed to be zero, slow waves propagate strictly along the cylinder’s axis and thus do not perturb the guiding field. For non-zero $\beta$, slow waves acquire a small perpendicular component (i.e., the $v_{r1}$ variable) and hence become locally oblique and cause perturbations of the cylinder’s magnetic field. This effect is manifested by the dependence of the wave speed upon the magnetic field (see the expression for $c_\mathrm{T}$). In particular, it makes slow waves an important tool for seismological diagnostics of the magnetic field in coronal loops \citep[e.g.][]{2007ApJ...656..598W}. Moreover, the oblique nature of slow waves is the unique feature that makes them sensitive to the dependence of the coronal heating function on magnetic field, considered in this work.}

The right-hand side of energy equation (\ref{eq:energy}) accounts for the field-aligned thermal conduction with the coefficient $\kappa_\parallel = 10^{-11}T_0^{5/2}\,\mathrm{W\,m}^{-1}\,\mathrm{K}^{-1}$ taken in the \textit{Spitzer} form in this work, and wave-induced perturbations of the coronal heat-loss function $Q(\rho,T,B)$. The latter is constituted by the balance between energy losses through the optically thin radiation ${\cal L}(\rho,T)$ and energy gains through an unspecified heating mechanism ${\cal H}(\rho, T, B)$, so that $Q(\rho,T,B) = {\cal L}(\rho,T) - {\cal H}(\rho, T, B)$. In the equilibrium, the coronal part of the loop is considered uniform with a constant temperature $T_0$, density $\rho_0$, and magnetic field $B_0$, providing $Q_0(\rho_0, T_0, B_0) = 0$ and no thermal conduction \citep[cf.][]{2022JApA...43...40K}.
$Q_\rho$, $Q_T$, and $Q_B$ on the right-hand side of energy equation (\ref{eq:energy}) represent coefficients in the Taylor expansion of the perturbed coronal heat-loss function in the vicinity of the equilibrium, i.e.
\begin{equation}
    Q(\rho,T,B) \approx Q_\rho\rho_1 + Q_T T_1 + Q_B B_1,
\end{equation}
with $Q_\rho = \partial Q/\partial\rho$, $Q_T = \partial Q/\partial T$, $Q_B = \partial Q/\partial B$, all taken at $T_0$, $\rho_0$, and $B_0$.

Assuming a harmonic form $\propto e^{i(kz - \omega t)}$ with the angular frequency $\omega$ and wavenumber $k$ for perturbations of all plasma parameters, Eqs.~(\ref{eq:continuity})--(\ref{eq:induction}) can be reduced to the following third-order polynomial dispersion relation \citep{2021A&A...646A.155D},
\begin{equation}
    \omega^3 + \Delta_2(k)\omega^2 + \Delta_1(k)\omega + \Delta_0(k) = 0,\label{eq:dispersion}
\end{equation}
with the coefficients
\begin{align}
    &\Delta_0(k) = -\frac{i}{C_\mathrm{V}(1+\gamma\beta/2)}\frac{k_\mathrm{B}T_0}{m}\left(Q_T-\frac{\rho_0}{T_0}Q_\rho+\frac{\kappa_\parallel k^2}{\rho_0}\right)k^2,\nonumber\\
    &\Delta_1(k) = -c_\mathrm{T}^2k^2,\nonumber\\
    &\Delta_2(k) = \frac{i}{C_\mathrm{V}(1+\gamma\beta/2)}\left[Q_T\left(1+\frac{\beta}{2}\right)-\frac{\rho_0}{T_0}\frac{\beta}{2}Q_\rho-\frac{B_0}{T_0}\frac{\beta}{2}Q_B + \frac{\kappa_\parallel k^2}{\rho_0}\left(1+\frac{\beta}{2}\right)\right].\nonumber
\end{align}
For $\beta \to 0$, Eq.~(\ref{eq:dispersion}) reduces to the dispersion relation considered in \citep{2019A&A...628A.133K}.
Being of the third-order in $\omega$, Eq.~(\ref{eq:dispersion}) describes the dynamics of two propagating slow magnetoacoustic modes and one non-propagating entropy mode. In other words, taking $\omega = \omega_\mathrm{R} + i\omega_\mathrm{I}$, gives us $\omega_\mathrm{R}\ne0$ and $\omega_\mathrm{I}\ne0$ for slow modes, and $\omega_\mathrm{R}=0$ and $\omega_\mathrm{I}\ne0$ for the entropy mode. Moreover, the amplitudes of slow and entropy modes exponentially grow (i.e. become unstable or overstable) for $\omega_\mathrm{I} > 0$.

To calculate the heat-loss function derivatives $Q_T$, $Q_\rho$, and $Q_B$, we use CHIANTI atomic database for modelling the optically thin radiative losses ${\cal L}(\rho,T)$ \citep{2021ApJ...909...38D}, and parametrise the unknown coronal heating function as ${\cal H}(\rho, T, B) \propto \rho^a T^b B^c$ \citep[see e.g.][]{1965ApJ...142..531F, 1978ApJ...220..643R, 1988SoPh..117...51D, 1992PPCF...34..411H, 1993ApJ...415..335I, 2006A&A...460..573C, 2022SoPh..297..144I}, with the power-law indices $(a,b,c)$ treated as free parameters in this work. The heating function is assumed to be uniform in space and its time dependence is omitted, assuming the apparent intermittent nature of coronal heating with time scales usually shorter than a minute \citep[e.g.][]{2014Sci...346B.315T, 2016ApJ...816...12T, 2019ApJ...884..131R, 2019ARA&A..57..157C} is effectively averaged over the slow wave oscillation period (from several minutes to a few tens of minutes).

Thus, the model described is similar to that considered in a series of previous works \citep{2017ApJ...849...62N, 2019PhPl...26h2113Z, 2019A&A...628A.133K, 2021A&A...646A.155D, 2021SoPh..296..105P, 2021PPCF...63l4008K, 2021SoPh..296..122B, 2022SoPh..297....5P, 2022MNRAS.514L..51K}, with a possible dependence of the coronal heating function on the magnetic field. The effect of this dependence on the wave stability is taken into account through the power-index $c$.

\section{Stabilisation by thermal conduction}
\label{sec:thermal_cond}
In a low-$\beta$ plasma, the dynamics of slow waves has been shown to be almost insensitive to the power-law index $c$ \citep{2021A&A...646A.155D, 2021PPCF...63l4008K}, as the waves produce negligible perturbations of the loop's magnetic field in this regime (also known as the infinite magnetic field approximation). For $\beta=0$, the stability of slow and entropy modes described by Eq.~(\ref{eq:dispersion}) is determined by the following conditions for $a$ and $b$ \citep{2020A&A...644A..33K},
\begin{align}
	&b < -\frac{a}{\gamma-1} + \frac{T_0}{{\cal L}_0}\frac{\partial {{\cal L}_0}}{\partial T}+ \frac{\kappa_\parallel k^2}{\rho_0}\frac{T_0}{{\cal L}_0} +\frac{1}{\gamma-1},\label{eq:instab_cond_ac_ab}\\
	&b < a + \frac{T_0}{{\cal L}_0}\frac{\partial {{\cal L}_0}}{\partial T} + \frac{\kappa_\parallel k^2}{\rho_0}\frac{T_0}{{\cal L}_0} - 1,  \label{eq:instab_cond_th_ab}
\end{align}
respectively.

In \cite{2020A&A...644A..33K}, the thermodynamic stability of the corona to slow and entropy modes was considered, and seismological constraints on the heating function parameters $a$ and $b$ were obtained for $\kappa_\parallel \to 0$ (or $k \to 0$), i.e. assuming the conductive term in Eqs.~(\ref{eq:instab_cond_ac_ab})--(\ref{eq:instab_cond_th_ab}) has no effect on the wave dynamics. In \cite{2021PPCF...63l4008K}, it was shown analytically that in the presence of thermal conduction, slow and entropy modes remain stable to perturbations shorter than the corresponding acoustic and thermal Field's lengths,
\begin{align}
	&\lambda_\mathrm{F}^\mathrm{thermal}=2\pi\sqrt{\frac{\kappa_\parallel T_0}{\rho_0{\cal L}_0\left[1-a+b-\displaystyle\frac{T_0}{{\cal L}_0}\frac{\partial {\cal L}_0}{\partial T}\right]}}\mathrm{\ \ } ,\label{eq:lambda_thermal}\\
	&\lambda_\mathrm{F}^\mathrm{acoustic}=2\pi\sqrt{\frac{\kappa_\parallel T_0}{\rho_0{\cal L}_0\left[\displaystyle\frac{a-1}{\gamma-1}+b-\displaystyle\frac{T_0}{{\cal L}_0}\frac{\partial {\cal L}_0}{\partial T}\right]}}\mathrm{\ \ }  ,\label{eq:lambda_acoustic}
\end{align}
but the plasma diagnostic potential of those conditions has not been explored.

\begin{figure}
	\begin{center}
		\includegraphics[width=0.49\textwidth]{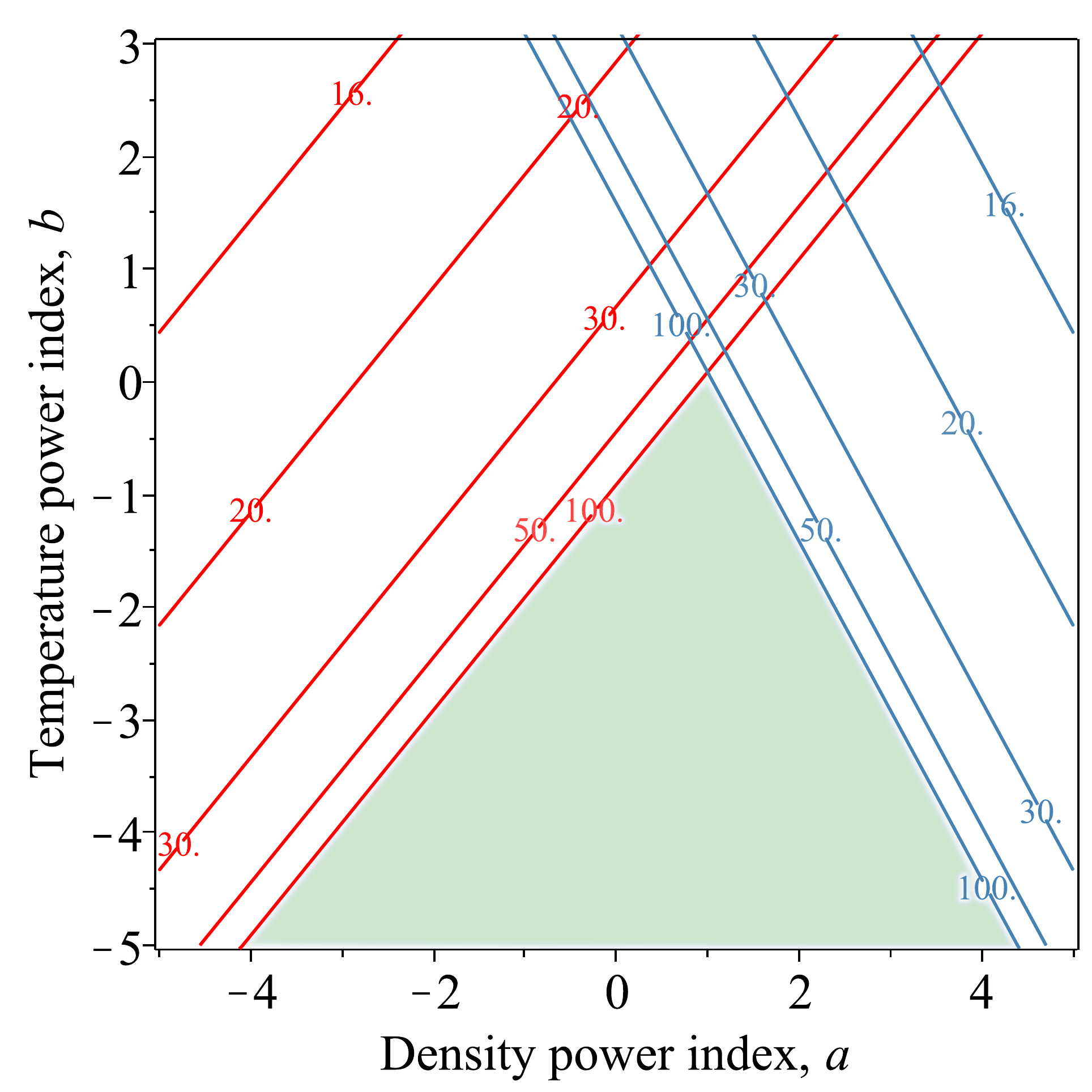}
		\includegraphics[width=0.49\textwidth]{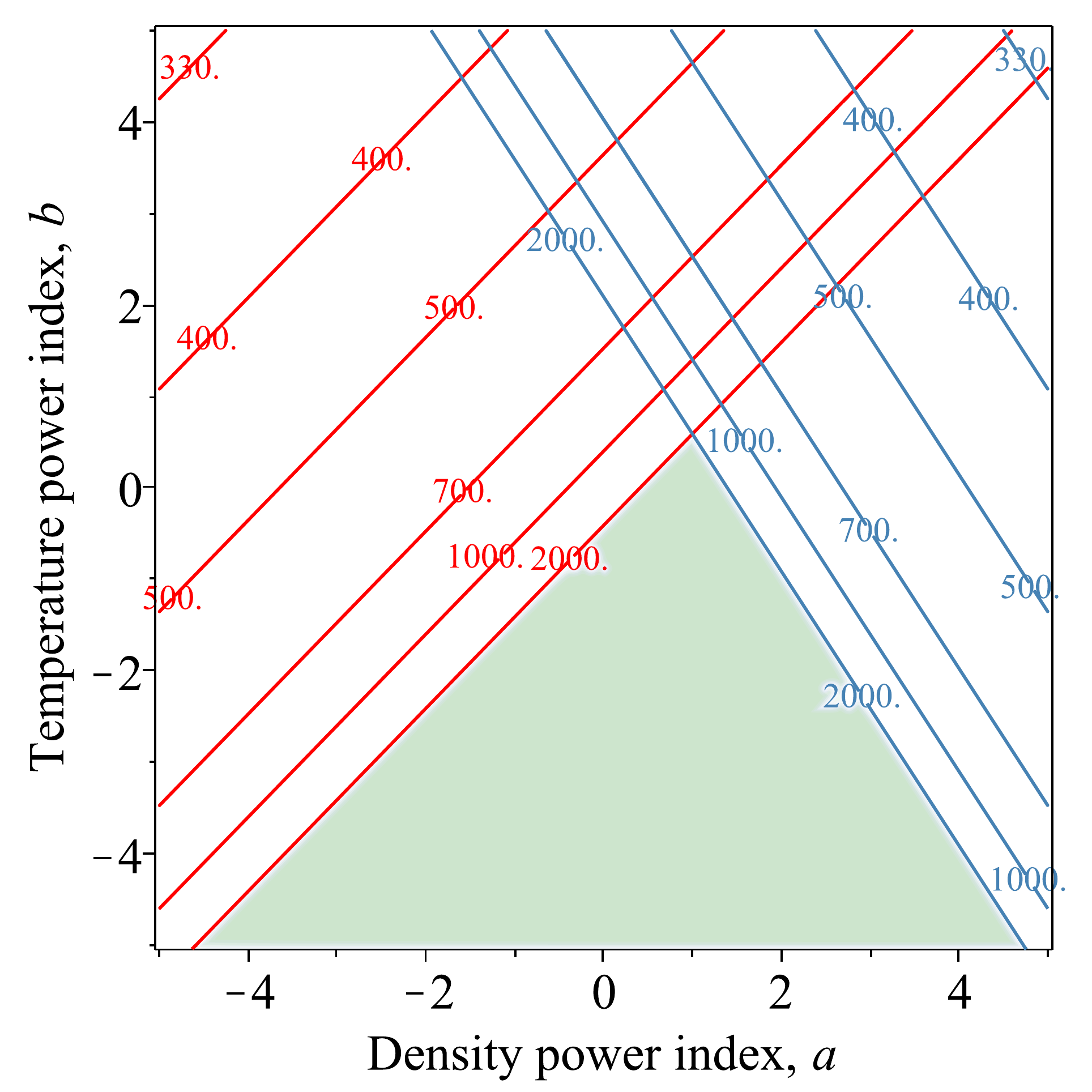}
	\end{center}
	\caption{Thermal (red) and acoustic (blue) Field's lengths $\lambda_\mathrm{F}^\mathrm{thermal}$ (\ref{eq:lambda_thermal}) and $\lambda_\mathrm{F}^\mathrm{acoustic}$ (\ref{eq:lambda_acoustic}) in Mm, respectively, evaluated for $\rho_0=3\times10^{-12}$\,kg\,m$^{-3}$ and $T_0=1$\,MK (left, typical for quiescent coronal loops hosting propagating slow waves), and $\rho_0=6\times10^{-12}$\,kg\,m$^{-3}$ and $T_0=6$\,MK (right, typical for hot active region loops hosting standing slow magnetoacoustic oscillations, also known as \lq\lq SUMER\rq\rq\ oscillations). In the green-shaded regions, both $\lambda_\mathrm{F}^\mathrm{thermal}$ and $\lambda_\mathrm{F}^\mathrm{acoustic}$ tend to infinity, i.e. the effect of the field-aligned thermal conduction on the wave stability vanishes.
	}
	\label{fig:1}
\end{figure}

In this work, we begin the analysis with evaluating $\lambda_\mathrm{F}^\mathrm{thermal}$ (\ref{eq:lambda_thermal}) and $\lambda_\mathrm{F}^\mathrm{acoustic}$ (\ref{eq:lambda_acoustic}) for plasma conditions typical for coronal structures hosting propagating slow waves and standing, SUMER-type slow oscillations. The obtained values of $\lambda_\mathrm{F}^\mathrm{thermal}$ and $\lambda_\mathrm{F}^\mathrm{acoustic}$  are shown in Fig.~\ref{fig:1} for various combination of the heating model parameters $(a,b)$. Thus, in quiescent coronal loops with typical temperature $T_0 = 1$\,MK and typical density $\rho_0 = 3\times 10^{-12}$\,kg\,m$^{-3}$ (the left panel in Fig.~\ref{fig:1}), the parametric domain of stability of the corona to slow and entropy modes in the $(a,b)$-plane broadens by thermal conduction, but this broadening is rather minor for typically observed wavelengths of propagating slow waves in the corona (50--100\,Mm). For the wavelengths longer than 100\,Mm, the effect of thermal conduction vanishes and the $(a,b)$-domain of stability reduces to that considered in \cite{2020A&A...644A..33K}.
In the right panel of Fig.~\ref{fig:1}, we show $\lambda_\mathrm{F}^\mathrm{thermal}$ (\ref{eq:lambda_thermal}) and $\lambda_\mathrm{F}^\mathrm{acoustic}$ (\ref{eq:lambda_acoustic}) and the corresponding heating function parameters $(a,b)$ for plasma conditions typical for SUMER-type oscillations in hot and dense flaring loops with  $T_0 = 6$\,MK and $\rho_0 = 6\times 10^{-12}$\,kg\,m$^{-3}$. In contrast to propagating waves in quiescent loops, slow and entropy modes are stable in the loops hosting SUMER oscillations, in a much broader interval of $(a,b)$. In particular, for typically observed wavelengths of SUMER oscillations of about 300\,Mm (prescribed by the loop length), slow and entropy modes remain stable throughout the entire domain of $(a,b)$ considered. For thermal conduction to become unimportant in this case, the oscillation wavelengths should be about a few thousand Mm which is unrealistic in the solar corona.

\section{Effect of finite plasma-$\beta$}
\label{sec:beta}

For a finite plasma-$\beta$, the condition for the slow mode stability (\ref{eq:instab_cond_ac_ab}) generalises to
\begin{equation}
	b < -\frac{a}{\gamma-1} + \frac{T_0}{{\cal L}_0}\frac{\partial {{\cal L}_0}}{\partial T}+ \frac{\kappa_\parallel k^2}{\rho_0}\frac{T_0}{{\cal L}_0} +\frac{1}{\gamma-1} + \frac{\beta}{2}\frac{\gamma}{\gamma - 1}c,\label{eq:instab_cond_ac_abc}
\end{equation}
which is now determined by three heating function power-indices $(a,b,c)$. It can be readily derived either in a weak or strong non-adiabaticity limit demanding $\omega_\mathrm{I} = 0$ \citep[see e.g. Eqs.~(19) and (27) in][]{2021A&A...646A.155D}. For the entropy mode, obtaining a similar condition for a finite plasma-$\beta$ is less obvious \citep[see, e.g.,][]{2021SoPh..296...96Z}. Hence, we solve dispersion relation (\ref{eq:dispersion}) numerically for $\omega$ and identify values of the heating parameters $(a,b,c)$ for which $\omega_\mathrm{I} = 0$ for both slow and entropy modes. The \textit{root\_scalar} function, from the \textit{Scipy}'s Optimization and Root Finding library, was employed with the \lq\lq Newton\rq\rq\ method. This solved Eq.~(\ref{eq:dispersion}) by implementing the Newton--Raphson method, which can work in the complex domain and converges quickly, given a sufficiently accurate initial guess for $\omega$. As the initial guess for the slow mode, we used approximate solutions for $\omega_\mathrm{R}$ and $\omega_\mathrm{I}$ derived using the first-order perturbation theory in the limit of weak non-adiabaticity \citep[see Eqs.~(18) and (19) in][]{2021A&A...646A.155D}. For the entropy mode, $\omega_\mathrm{R} = 0$ was taken as the initial guess.

\begin{figure}
	\begin{center}
		\includegraphics[width=0.49\textwidth]{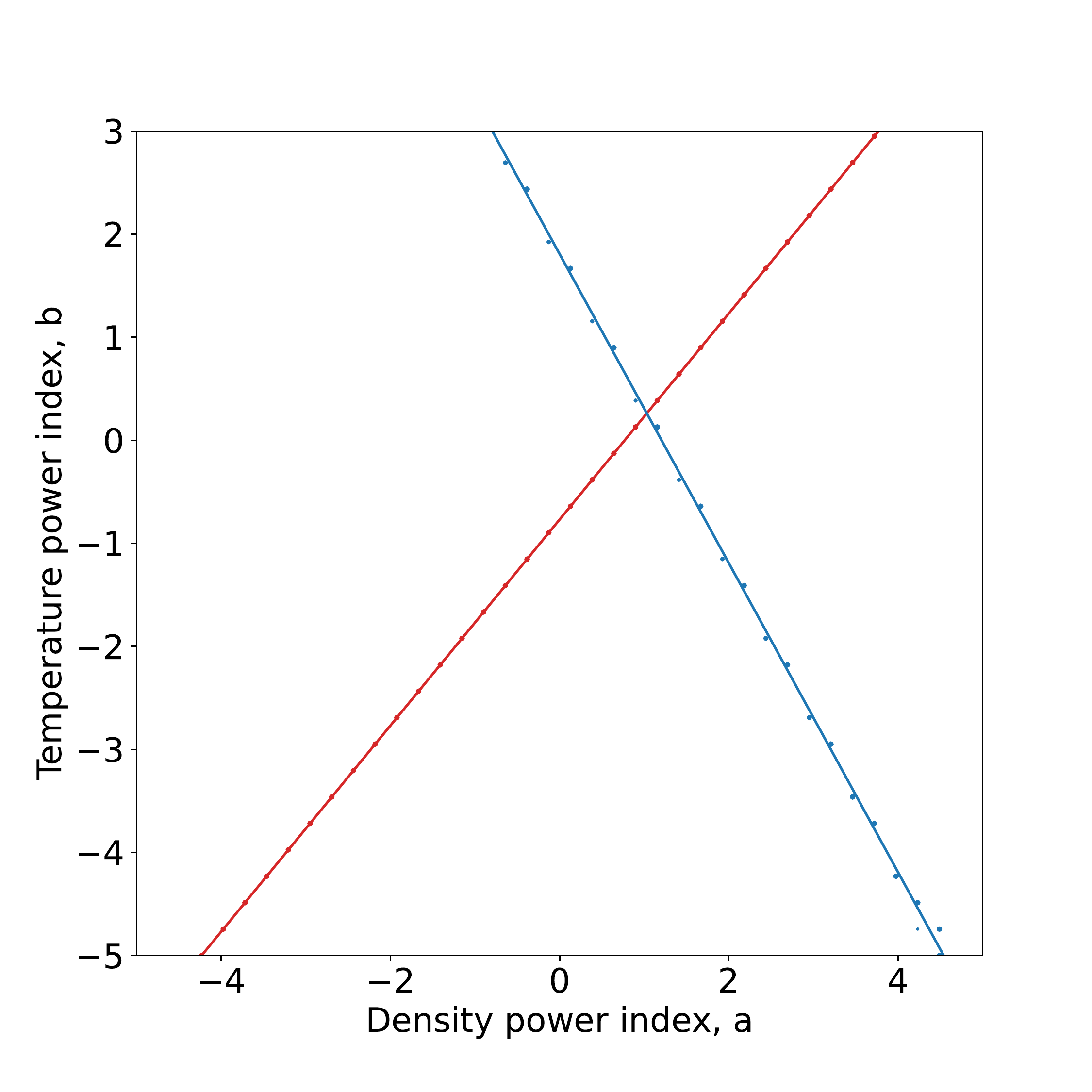}
		\includegraphics[width=0.49\textwidth]{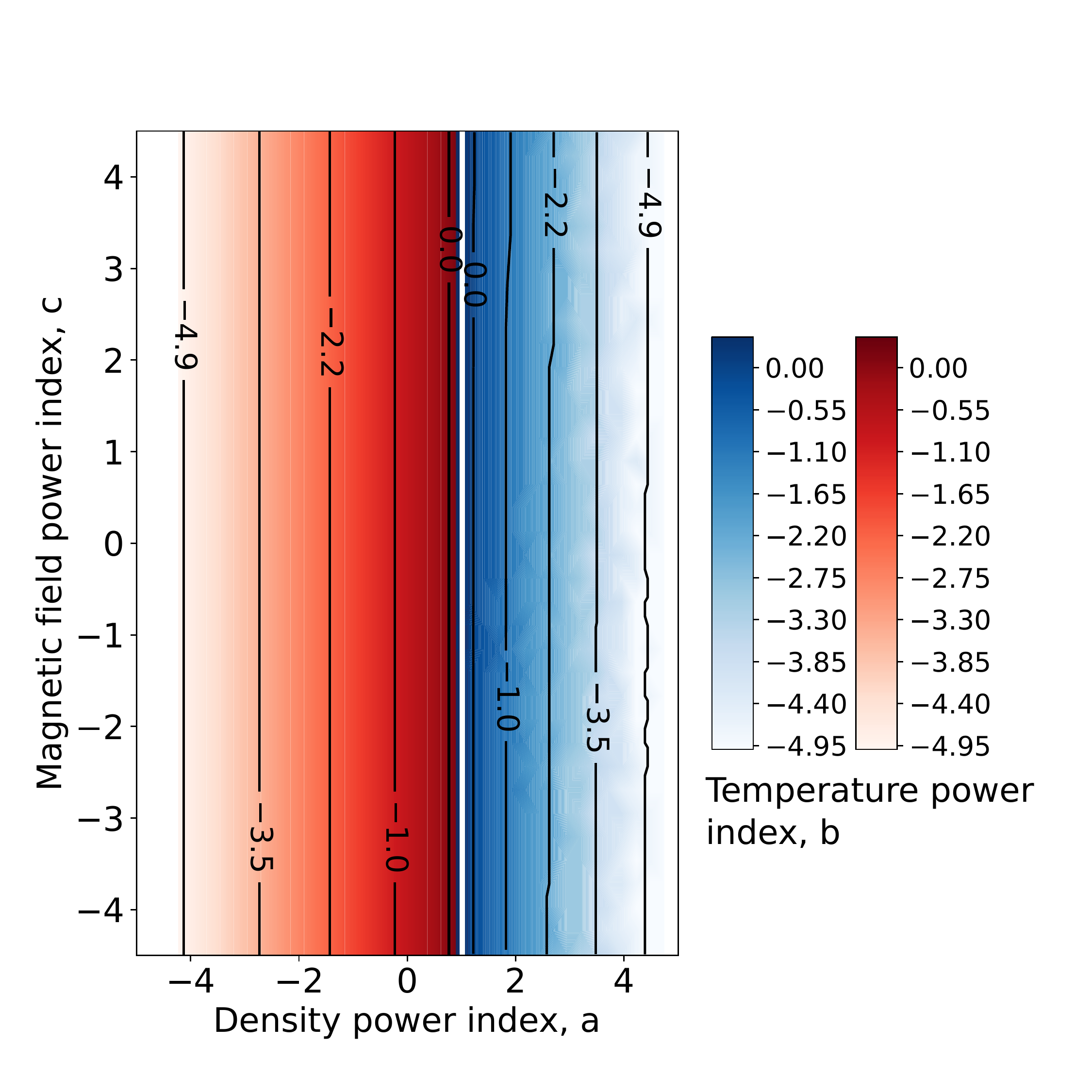}
		\includegraphics[width=0.49\textwidth]{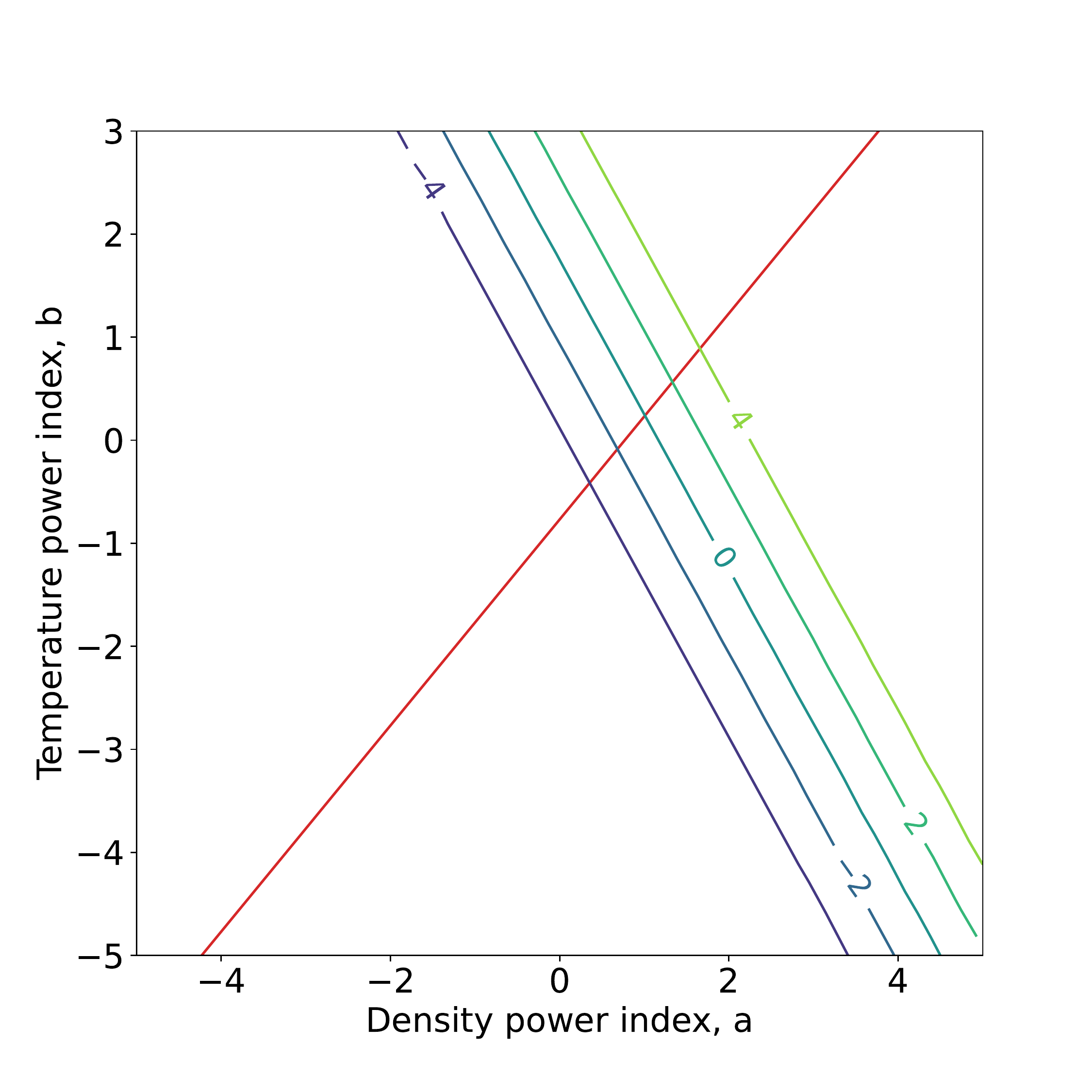}
		\includegraphics[width=0.49\textwidth]{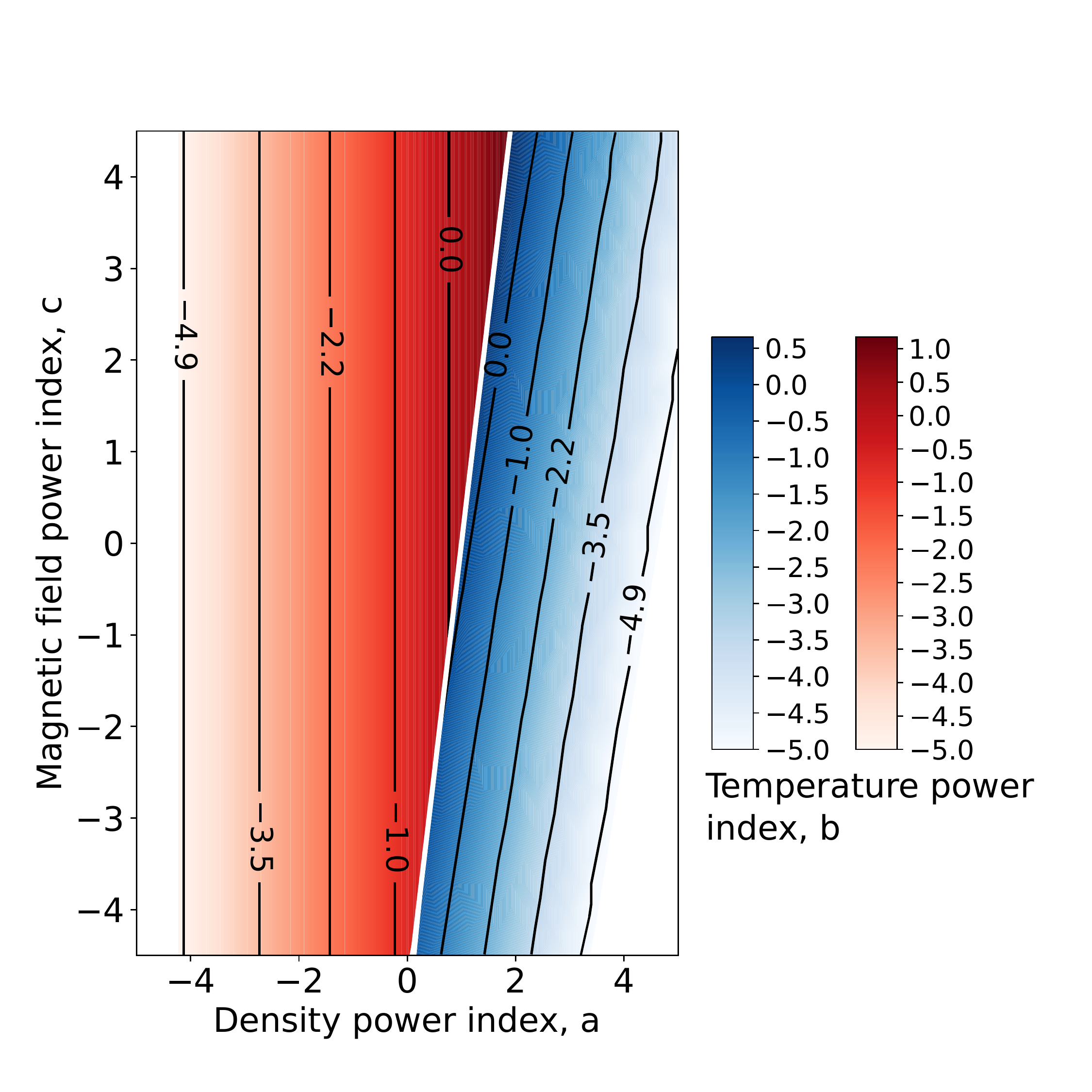}
	\end{center}
	\caption{Values of the power-indices $(a,b,c)$ in the parametrisation of the coronal heating function ${\cal H}(\rho, T, B)\propto \rho^aT^bB^c$, below which the coronal plasma is stable to slow (gradient of blue) and entropy (gradient of red) wave modes. The stability conditions are obtained from the numerical solution of dispersion relation (\ref{eq:dispersion}) for plasma parameters typical of quiescent coronal loops with low (top) and finite (bottom) values of the plasma parameter $\beta$, hosting propagating slow waves: $\rho_0=3\times10^{-12}$\,kg\,m$^{-3}$; $T_0=1$\,MK providing the adiabatic sound speed $C_\mathrm{s}\approx152\times\sqrt{T_0[\mathrm{MK}]}\approx 152$\,km\,s$^{-1}$; $B_0=20$\,G (top) and $B_0=4$\,G (bottom) providing plasma-$\beta\approx0.01$ and $\beta\approx0.3$, respectively; $\lambda = 70$\,Mm (providing the adiabatic slow wave period $\lambda/C_\mathrm{s}\approx 7$\,min). The red and blue dots in the top left panel illustrate the grid size used in our numerical solution. The contour labels in the bottom left panel and in the right panels indicate values of the power indices $c$ and $b$, respectively. The white lines in the right panels indicate the intersection of the $abc$-planes determined by the slow and entropy wave stability conditions.
	}
	\label{fig:2}
\end{figure}

\begin{figure}
	\begin{center}
		\includegraphics[width=0.49\textwidth]{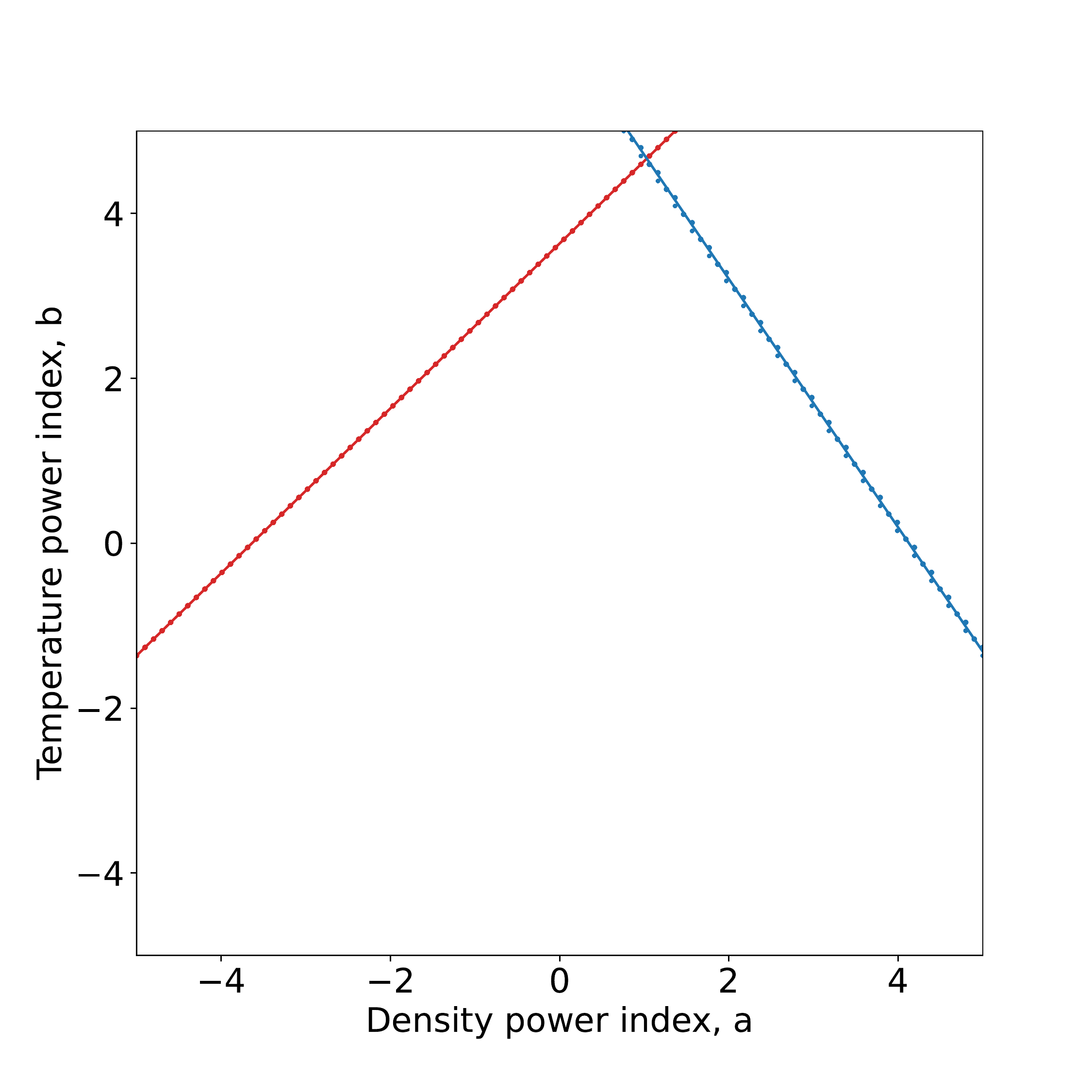}
		\includegraphics[width=0.49\textwidth]{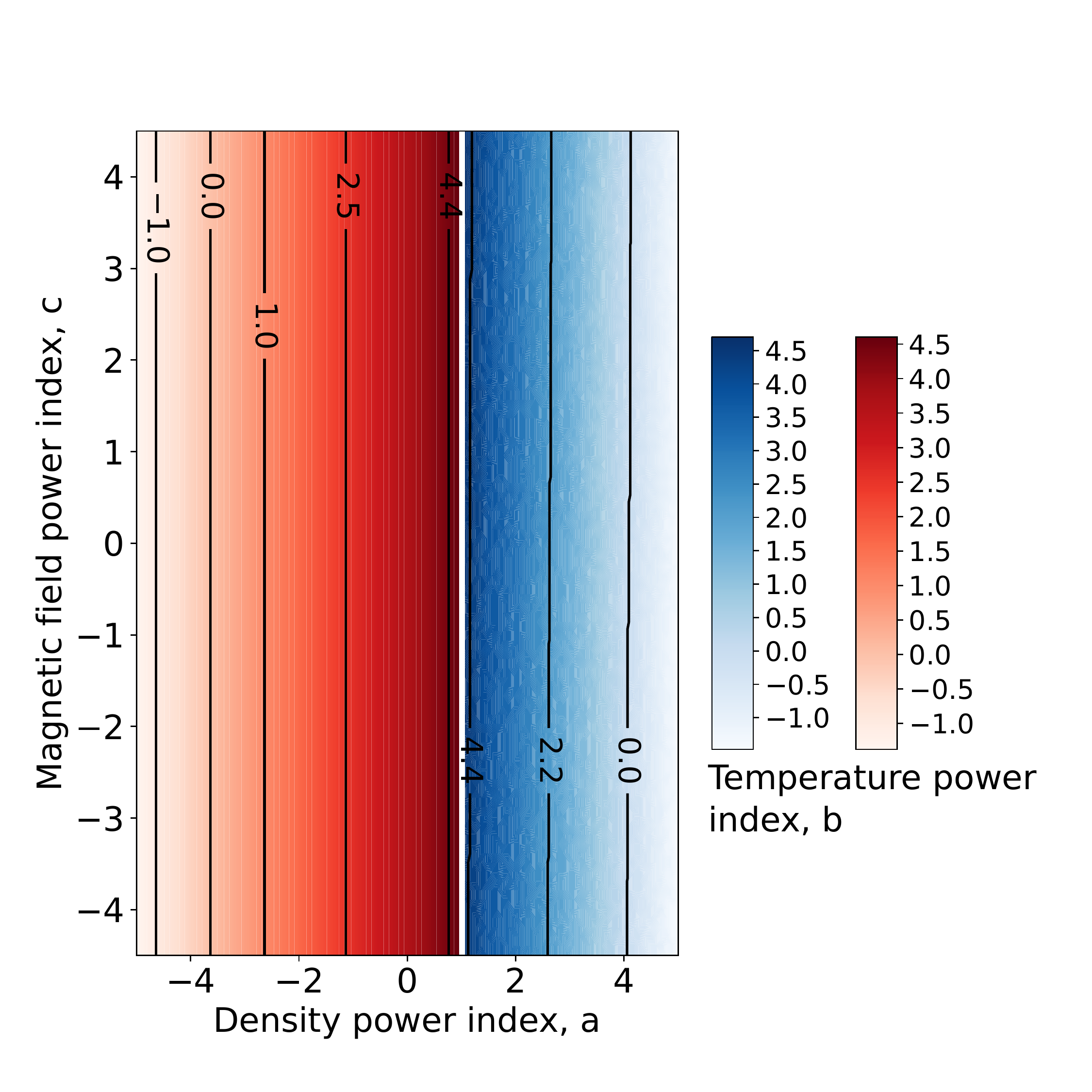}
		\includegraphics[width=0.49\textwidth]{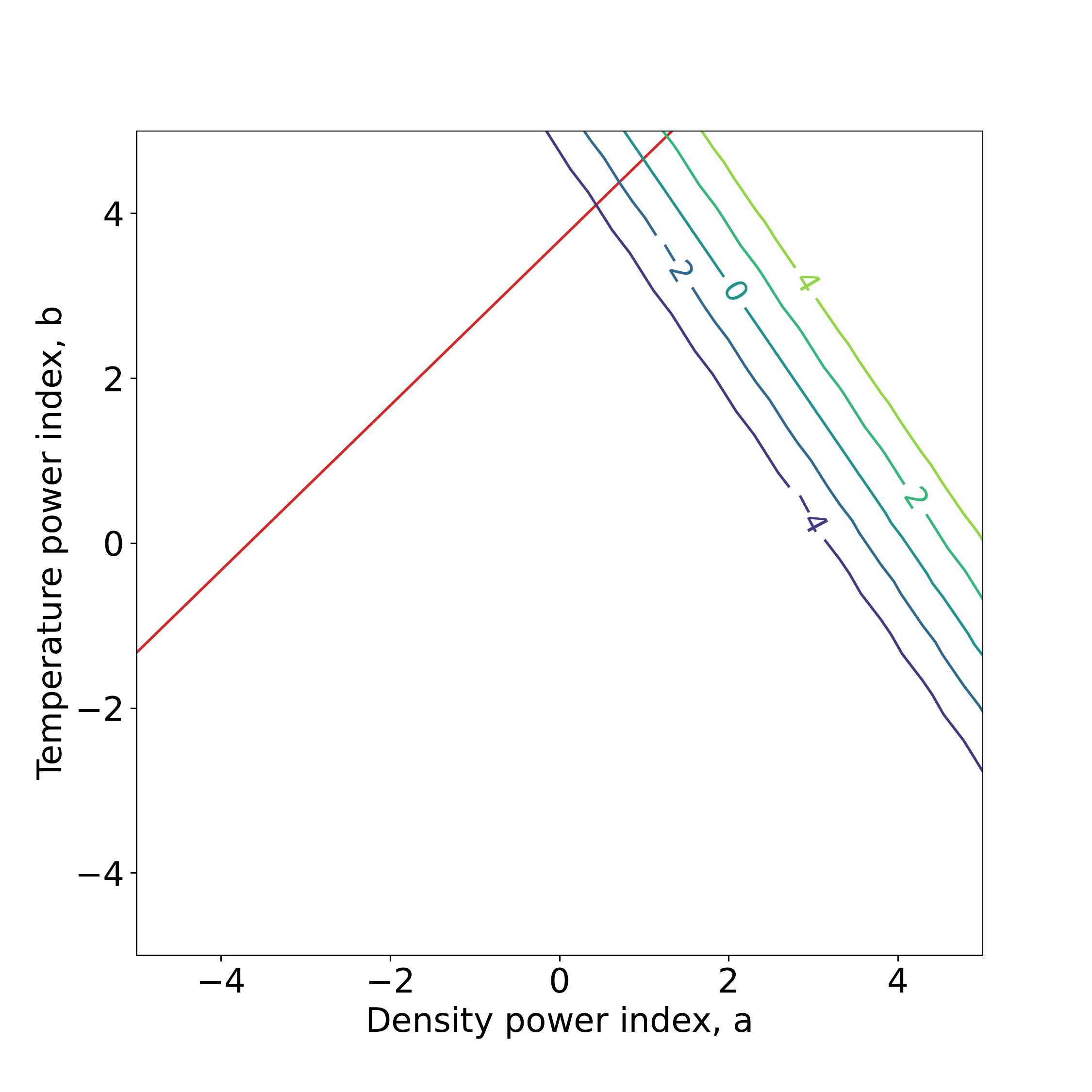}
		\includegraphics[width=0.49\textwidth]{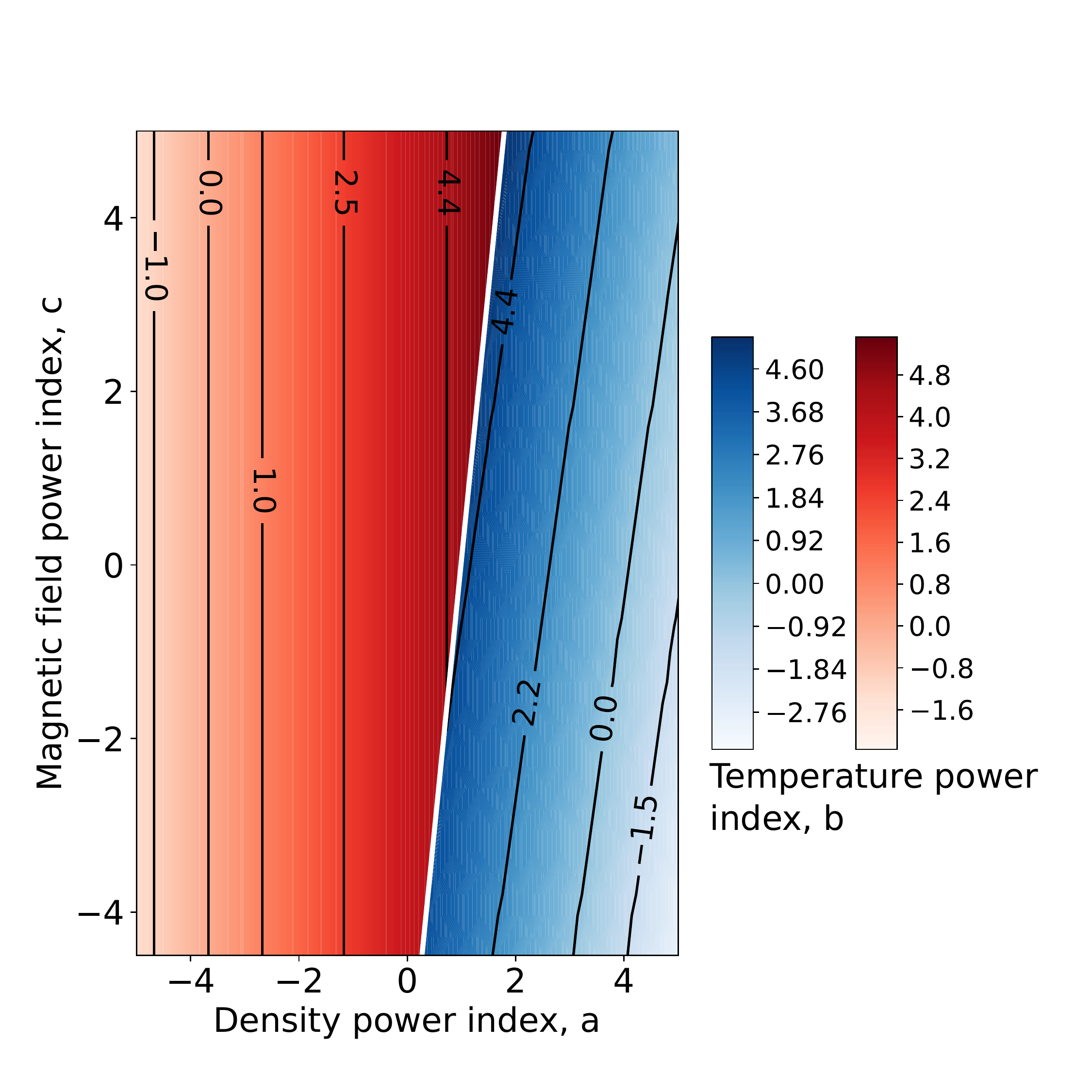}
	\end{center}
	\caption{The same as in Fig.~\ref{fig:2} but for SUMER-type loops hosting standing slow oscillations with low (top) and finite (bottom) values of the plasma parameter $\beta$: $\rho_0=6\times10^{-12}$\,kg\,m$^{-3}$; $T_0=6$\,MK providing the adiabatic sound speed $C_\mathrm{s}\approx152\times\sqrt{T_0[\mathrm{MK}]}\approx 372$\,km\,s$^{-1}$; $B_0=80$\,G (top) and $B_0=15$\,G (bottom) providing plasma-$\beta\approx0.01$ and $\beta\approx0.3$, respectively; $\lambda = 500$\,Mm (providing the adiabatic slow wave period $\lambda/C_\mathrm{s}\approx 22$\,min).
	}
	\label{fig:3}
\end{figure}

Figures \ref{fig:2} and \ref{fig:3} show domains of the slow and entropy mode stability in the $(a,b,c)$ parametric space, obtained numerically for the above-mentioned plasma density and temperature in coronal loops hosting propagating slow waves and SUMER-type oscillations, and with plasma-$\beta$ of about 0.01 and 0.3. In the low-$\beta$ regime (top panels in Figs. \ref{fig:2} and \ref{fig:3}), the stability of the corona to slow and entropy modes is almost independent of the value of the magnetic field power-index $c$ in the power-law parametrisation of the unknown coronal heating function. This is consistent with the outcomes of the zero-$\beta$ analysis described in Sec.~\ref{sec:thermal_cond}. In the finite-$\beta$ case (bottom panels in Figs. \ref{fig:2} and \ref{fig:3}), the $(a,b,c)$ domain determining the stability of the slow mode gets modified, so that higher values of the parameter $c$ result in broader intervals of $a$ and $b$ for which slow waves remain stable. The latter is consistent with the analytical prediction by Eq.~(\ref{eq:instab_cond_ac_abc}). In contrast, the stability condition for the entropy mode is found to be insensitive to the parameter $c$, with density and temperature power-indices $(a,b)$ and thermal conduction playing the major role in terms of the considered model. The described tendencies in the dependence of the coronal plasma stability to slow and entropy modes on the coronal heating function parametrisation with the local plasma density, temperature, and magnetic field are found to be qualitatively similar for both propagating slow waves in quiescent coronal loops and standing, SUMER-type oscillations in hot loops.

\section{Discussion and conclusions}
\label{sec:disc}

We examined the stability of slow magnetoacoustic and entropy waves in the solar corona to the wave-induced perturbation of the local thermal equilibrium between energy losses through optically thin radiation and energy gains from an unspecified coronal heating process, i.e., the effect of wave-induced thermal misbalance. Using the first-order thin flux tube approximation, we generalised the stability analysis performed in  previous works for regimes with finite \textit{Spitzer} thermal conduction along the field and finite plasma parameter $\beta$, taking the possible dependence of the unknown coronal heating function on the local value of the magnetic field in its generic parametrisation ${\cal H}(\rho, T, B)\propto \rho^aT^bB^c$ into account. We applied the generalised theory to physical conditions typical for quiescent coronal plasma loops and hot active region loops, in which propagating slow waves and standing SUMER oscillations are respectively observed, treating the power-indices $(a,b,c)$ and the plasma-$\beta$ as free parameters. The results of our study can be summarised as:
\begin{itemize}
    \item The field-aligned thermal conduction tends to stabilise both slow and entropy modes, thus effectively broadening the parametric domain in the $(a,b,c)$ space, within which the coronal plasma remains stable. The efficiency of this stabilisation by thermal conduction, however, strongly depends on the equilibrium plasma conditions and perturbation wavelength, determined by the thermal and acoustic Field's lengths given by Eqs.~ (\ref{eq:lambda_thermal}) and (\ref{eq:lambda_acoustic}), respectively. While for slow waves in quiescent loops this effect is found to be rather minor, hot loops are shown to be predominantly stable to slow and entropy modes due to highly effective thermal conduction. In other words, the considered functional form of the heating model and the field-aligned \textit{Spitzer} conductivity cannot account for the development of slow magnetoacoustic overstability and/or rapid coronal condensations, associated with isentropic and isobaric thermal instabilities in hot active region loops \citep[see e.g.][for the most recent review]{2022FrASS...920116A}.

    \item The stability of the entropy mode is found to be insensitive to the dependence of the coronal heating function on the local magnetic field, i.e. the parameter $c$, and plasma-$\beta$. In other words, this result suggests that catastrophic cooling and condensations of the coronal plasma, caused by the entropy mode instability, are independent of the magnetic properties of host active regions and are fully driven by the loss of balance between optically thin radiation, field-aligned conduction, and heating, which is consistent with numerical findings of, e.g., \citep{2016ApJ...823...22X}. On the other hand, there could be an indirect link through the local plasma heating by magnetic reconnection \citep[see e.g.][]{2018ApJ...864L...4L}, which is out of the scope of our work.

    \item The stability of slow magnetoacoustic waves, in contrast, is found to depend on the product of two magnetic field parameters, $\beta c$. Thus, in finite-$\beta$ plasma, one may expect to probe the functional dependence of the coronal heating function on the magnetic field, i.e. the parameter $c$, with the theory presented here. This result effectively extends the seismological diagnostics of the coronal heating function developed in \cite{2020A&A...644A..33K}. In particular, it allows us to consider both existing AC and DC theories of coronal heating \citep[see, e.g., Table 5 in][]{2000ApJ...530..999M} for comparison and validation.
    {For example, following \citet{1978ApJ...220..643R}, \citet{1993ApJ...415..335I} considered coronal heating mechanisms by electric current dissipation, mode conversion, and anomalous conduction damping of Alfv\'en waves  in a similar power-law form.}
    Together with the diagnostic potential of microwave observations \citep[see][]{2021ApJ...909...89F}, such a seismological technique offers a unique opportunity to constrain the link between the coronal magnetic field and thermodynamic properties of the corona, which is not directly available in extreme ultraviolet or soft X-ray observations traditionally used for coronal heating studies.
\end{itemize}

Thus, the effect of the back-reaction of wave-induced thermal misbalance is an important element of the coronal dynamics and structure. Accounting for this effect determines conditions for stability of slow magnetoacoustic and entropy perturbations in the corona. As the next step, the comparison of the theoretical results obtained in this work with observed properties of coronal slow waves will allow us to effectively constrain the modelling of physical mechanisms responsible for the formation and existence of hot plasmas in the solar and stellar coronae. 
%%%%%%%%%%%%%%%%%%%%%%%%%%%%%%%%%%%%%%%%%%
\vspace{6pt} 

%%%%%%%%%%%%%%%%%%%%%%%%%%%%%%%%%%%%%%%%%%
%% optional
%\supplementary{The following supporting information can be downloaded at:  \linksupplementary{s1}, Figure S1: title; Table S1: title; Video S1: title.}


%%%%%%%%%%%%%%%%%%%%%%%%%%%%%%%%%%%%%%%%%%
\authorcontributions{DYK was responsible for the formulation of the problem, choice of the model and analytical techniques, producing figures, and writing up Sections~\ref{sec:model}, \ref{sec:thermal_cond}, \ref{sec:beta}, and \ref{sec:disc}. VMN wrote Section~\ref{sec:intro}, contributed to writing Secction~\ref{sec:disc}, and participated in the discussion and interpretation of the obtained results. JBF derived Eq.~(\ref{eq:instab_cond_ac_abc}), solved Eq.~(\ref{eq:dispersion}) numerically in Python, and contributed to writing Section~\ref{sec:beta}. All authors contributed to proofreading and editing the manuscript.}

\funding{DYK and VMN acknowledge support from the STFC Consolidated Grant ST/T000252/1. The work of JBF was supported by the Undergraduate Research Support Scheme from the University of Warwick.}

\dataavailability{Not applicable.} 

\acknowledgments{DYK and VMN are grateful to Marcel Goossens for inspiring discussions and continuous encouragement. 
The authors thank the International Online Team \lq\lq Effects of Coronal Heating/Cooling on MHD Waves\rq\rq\ for their interest and active debates.}

\conflictsofinterest{The authors declare no conflict of interest. The funders had no role in the design
of the study; in the collection, analyses, or interpretation of data; in the writing of the manuscript; or in the decision to publish the results.} 

%%%%%%%%%%%%%%%%%%%%%%%%%%%%%%%%%%%%%%%%%%
%% Optional
%\sampleavailability{Samples of the compounds ... are available from the authors.}

%% Only for journal Encyclopedia
%\entrylink{The Link to this entry published on the encyclopedia platform.}

%\abbreviations{Abbreviations}{
%The following abbreviations are used in this manuscript:\\

%\noindent 
%\begin{tabular}{@{}ll}
%MDPI & Multidisciplinary Digital Publishing Institute\\
%DOAJ & Directory of open access journals\\
%TLA & Three letter acronym\\
%LD & Linear dichroism
%\end{tabular}
%}

%%%%%%%%%%%%%%%%%%%%%%%%%%%%%%%%%%%%%%%%%%
\begin{adjustwidth}{-\extralength}{0cm}

\reftitle{References}

\end{adjustwidth}

\begin{thebibliography}{999}
	
	\bibitem[{Nakariakov} and {Kolotkov}(2020)]{2020ARA&A..58..441N}
	{Nakariakov}, V.M.; {Kolotkov}, D.Y.
	\newblock {Magnetohydrodynamic Waves in the Solar Corona}.
	\newblock {\em \araa} {\bf 2020}, {\em 58},~441--481.
	\newblock {\url{https://doi.org/10.1146/annurev-astro-032320-042940}}.
	
	\bibitem[{Van Doorsselaere} \em{et~al.}(2020){Van Doorsselaere}, {Srivastava},
	{Antolin}, {Magyar}, {Vasheghani Farahani}, {Tian}, {Kolotkov}, {Ofman},
	{Guo}, {Arregui}, {De Moortel}, and {Pascoe}]{2020SSRv..216..140V}
	{Van Doorsselaere}, T.; {Srivastava}, A.K.; {Antolin}, P.; {Magyar}, N.;
	{Vasheghani Farahani}, S.; {Tian}, H.; {Kolotkov}, D.; {Ofman}, L.; {Guo},
	M.; {Arregui}, I.;  et~al.
	\newblock {Coronal Heating by MHD Waves}.
	\newblock {\em \ssr} {\bf 2020}, {\em 216},~140,
	\href{http://xxx.lanl.gov/abs/2012.01371}{{\normalfont
			[arXiv:astro-ph.SR/2012.01371]}}.
	\newblock {\url{https://doi.org/10.1007/s11214-020-00770-y}}.
	
	\bibitem[{De Moortel} and {Nakariakov}(2012)]{2012RSPTA.370.3193D}
	{De Moortel}, I.; {Nakariakov}, V.M.
	\newblock {Magnetohydrodynamic waves and coronal seismology: an overview of
		recent results}.
	\newblock {\em Philosophical Transactions of the Royal Society of London Series
		A} {\bf 2012}, {\em 370},~3193--3216,
	\href{http://xxx.lanl.gov/abs/1202.1944}{{\normalfont
			[arXiv:astro-ph.SR/1202.1944]}}.
	\newblock {\url{https://doi.org/10.1098/rsta.2011.0640}}.
	
	\bibitem[{Banerjee} \em{et~al.}(2021){Banerjee}, {Krishna Prasad}, {Pant},
	{McLaughlin}, {Antolin}, {Magyar}, {Ofman}, {Tian}, {Van Doorsselaere}, {De
		Moortel}, and {Wang}]{2021SSRv..217...76B}
	{Banerjee}, D.; {Krishna Prasad}, S.; {Pant}, V.; {McLaughlin}, J.A.;
	{Antolin}, P.; {Magyar}, N.; {Ofman}, L.; {Tian}, H.; {Van Doorsselaere}, T.;
	{De Moortel}, I.;  et~al.
	\newblock {Magnetohydrodynamic Waves in Open Coronal Structures}.
	\newblock {\em \ssr} {\bf 2021}, {\em 217},~76,
	\href{http://xxx.lanl.gov/abs/2012.08802}{{\normalfont
			[arXiv:astro-ph.SR/2012.08802]}}.
	\newblock {\url{https://doi.org/10.1007/s11214-021-00849-0}}.
	
	\bibitem[{De Moortel}(2009)]{2009SSRv..149...65D}
	{De Moortel}, I.
	\newblock {Longitudinal Waves in Coronal Loops}.
	\newblock {\em \ssr} {\bf 2009}, {\em 149},~65--81.
	\newblock {\url{https://doi.org/10.1007/s11214-009-9526-5}}.
	
	\bibitem[{Wang} \em{et~al.}(2021){Wang}, {Ofman}, {Yuan}, {Reale}, {Kolotkov},
	and {Srivastava}]{2021SSRv..217...34W}
	{Wang}, T.; {Ofman}, L.; {Yuan}, D.; {Reale}, F.; {Kolotkov}, D.Y.;
	{Srivastava}, A.K.
	\newblock {Slow-Mode Magnetoacoustic Waves in Coronal Loops}.
	\newblock {\em \ssr} {\bf 2021}, {\em 217},~34,
	\href{http://xxx.lanl.gov/abs/2102.11376}{{\normalfont
			[arXiv:astro-ph.SR/2102.11376]}}.
	\newblock {\url{https://doi.org/10.1007/s11214-021-00811-0}}.
	
	\bibitem[{Nakariakov} \em{et~al.}(2019){Nakariakov}, {Kosak}, {Kolotkov},
	{Anfinogentov}, {Kumar}, and {Moon}]{2019ApJ...874L...1N}
	{Nakariakov}, V.M.; {Kosak}, M.K.; {Kolotkov}, D.Y.; {Anfinogentov}, S.A.;
	{Kumar}, P.; {Moon}, Y.J.
	\newblock {Properties of Slow Magnetoacoustic Oscillations of Solar Coronal
		Loops by Multi-instrumental Observations}.
	\newblock {\em \apjl} {\bf 2019}, {\em 874},~L1.
	\newblock {\url{https://doi.org/10.3847/2041-8213/ab0c9f}}.
	
	\bibitem[{Pant} \em{et~al.}(2017){Pant}, {Tiwari}, {Yuan}, and
	{Banerjee}]{2017ApJ...847L...5P}
	{Pant}, V.; {Tiwari}, A.; {Yuan}, D.; {Banerjee}, D.
	\newblock {First Imaging Observation of Standing Slow Wave in Coronal Fan
		Loops}.
	\newblock {\em \apjl} {\bf 2017}, {\em 847},~L5,
	\href{http://xxx.lanl.gov/abs/1708.06946}{{\normalfont
			[arXiv:astro-ph.SR/1708.06946]}}.
	\newblock {\url{https://doi.org/10.3847/2041-8213/aa880f}}.
	
	\bibitem[{Wang}(2011)]{2011SSRv..158..397W}
	{Wang}, T.
	\newblock {Standing Slow-Mode Waves in Hot Coronal Loops: Observations,
		Modeling, and Coronal Seismology}.
	\newblock {\em \ssr} {\bf 2011}, {\em 158},~397--419,
	\href{http://xxx.lanl.gov/abs/1011.2483}{{\normalfont
			[arXiv:astro-ph.SR/1011.2483]}}.
	\newblock {\url{https://doi.org/10.1007/s11214-010-9716-1}}.
	
	\bibitem[{Kumar} \em{et~al.}(2013){Kumar}, {Innes}, and
	{Inhester}]{2013ApJ...779L...7K}
	{Kumar}, P.; {Innes}, D.E.; {Inhester}, B.
	\newblock {Solar Dynamics Observatory/Atmospheric Imaging Assembly Observations
		of a Reflecting Longitudinal Wave in a Coronal Loop}.
	\newblock {\em \apjl} {\bf 2013}, {\em 779},~L7,
	\href{http://xxx.lanl.gov/abs/1409.3896}{{\normalfont
			[arXiv:astro-ph.SR/1409.3896]}}.
	\newblock {\url{https://doi.org/10.1088/2041-8205/779/1/L7}}.
	
	\bibitem[{Reale}(2016)]{2016ApJ...826L..20R}
	{Reale}, F.
	\newblock {Plasma Sloshing in Pulse-heated Solar and Stellar Coronal Loops}.
	\newblock {\em \apjl} {\bf 2016}, {\em 826},~L20,
	\href{http://xxx.lanl.gov/abs/1607.01329}{{\normalfont
			[arXiv:astro-ph.SR/1607.01329]}}.
	\newblock {\url{https://doi.org/10.3847/2041-8205/826/2/L20}}.
	
	\bibitem[{Krishna Prasad} and {Van Doorsselaere}(2021)]{2021ApJ...914...81K}
	{Krishna Prasad}, S.; {Van Doorsselaere}, T.
	\newblock {Compressive Oscillations in Hot Coronal Loops: Are Sloshing
		Oscillations and Standing Slow Waves Independent?}
	\newblock {\em \apj} {\bf 2021}, {\em 914},~81,
	\href{http://xxx.lanl.gov/abs/2104.12038}{{\normalfont
			[arXiv:astro-ph.SR/2104.12038]}}.
	\newblock {\url{https://doi.org/10.3847/1538-4357/abfb01}}.
	
	\bibitem[{Ofman} \em{et~al.}(1999){Ofman}, {Nakariakov}, and
	{DeForest}]{1999ApJ...514..441O}
	{Ofman}, L.; {Nakariakov}, V.M.; {DeForest}, C.E.
	\newblock {Slow Magnetosonic Waves in Coronal Plumes}.
	\newblock {\em \apj} {\bf 1999}, {\em 514},~441--447.
	\newblock {\url{https://doi.org/10.1086/306944}}.
	
	\bibitem[{Ofman} and {Wang}(2002)]{2002ApJ...580L..85O}
	{Ofman}, L.; {Wang}, T.
	\newblock {Hot Coronal Loop Oscillations Observed by SUMER: Slow Magnetosonic
		Wave Damping by Thermal Conduction}.
	\newblock {\em \apjl} {\bf 2002}, {\em 580},~L85--L88.
	\newblock {\url{https://doi.org/10.1086/345548}}.
	
	\bibitem[{Nakariakov} \em{et~al.}(2000){Nakariakov}, {Verwichte}, {Berghmans},
	and {Robbrecht}]{2000A&A...362.1151N}
	{Nakariakov}, V.M.; {Verwichte}, E.; {Berghmans}, D.; {Robbrecht}, E.
	\newblock {Slow magnetoacoustic waves in coronal loops}.
	\newblock {\em \aap} {\bf 2000}, {\em 362},~1151--1157.
	
	\bibitem[{Tsiklauri} and {Nakariakov}(2001)]{2001A&A...379.1106T}
	{Tsiklauri}, D.; {Nakariakov}, V.M.
	\newblock {Wide-spectrum slow magnetoacoustic waves in coronal loops}.
	\newblock {\em \aap} {\bf 2001}, {\em 379},~1106--1112,
	\href{http://xxx.lanl.gov/abs/astro-ph/0107579}{{\normalfont
			[arXiv:astro-ph/astro-ph/0107579]}}.
	\newblock {\url{https://doi.org/10.1051/0004-6361:20011378}}.
	
	\bibitem[{De Moortel} and {Hood}(2003)]{2003A&A...408..755D}
	{De Moortel}, I.; {Hood}, A.W.
	\newblock {The damping of slow MHD waves in solar coronal magnetic fields}.
	\newblock {\em \aap} {\bf 2003}, {\em 408},~755--765.
	\newblock {\url{https://doi.org/10.1051/0004-6361:20030984}}.
	
	\bibitem[{Selwa} \em{et~al.}(2005){Selwa}, {Murawski}, and
	{Solanki}]{2005A&A...436..701S}
	{Selwa}, M.; {Murawski}, K.; {Solanki}, S.K.
	\newblock {Excitation and damping of slow magnetosonic standing waves in a
		solar coronal loop}.
	\newblock {\em \aap} {\bf 2005}, {\em 436},~701--709.
	\newblock {\url{https://doi.org/10.1051/0004-6361:20042319}}.
	
	\bibitem[{Ruderman}(2013)]{2013A&A...553A..23R}
	{Ruderman}, M.S.
	\newblock {Nonlinear damped standing slow waves in hot coronal magnetic loops}.
	\newblock {\em \aap} {\bf 2013}, {\em 553},~A23.
	\newblock {\url{https://doi.org/10.1051/0004-6361/201321175}}.
	
	\bibitem[{Yuan} \em{et~al.}(2015){Yuan}, {Van Doorsselaere}, {Banerjee}, and
	{Antolin}]{2015ApJ...807...98Y}
	{Yuan}, D.; {Van Doorsselaere}, T.; {Banerjee}, D.; {Antolin}, P.
	\newblock {Forward Modeling of Standing Slow Modes in Flaring Coronal Loops}.
	\newblock {\em \apj} {\bf 2015}, {\em 807},~98,
	\href{http://xxx.lanl.gov/abs/1504.07475}{{\normalfont
			[arXiv:astro-ph.SR/1504.07475]}}.
	\newblock {\url{https://doi.org/10.1088/0004-637X/807/1/98}}.
	
	\bibitem[{Field}(1965)]{1965ApJ...142..531F}
	{Field}, G.B.
	\newblock {Thermal Instability.}
	\newblock {\em \apj} {\bf 1965}, {\em 142},~531.
	\newblock {\url{https://doi.org/10.1086/148317}}.
	
	\bibitem[{van der Linden} and {Goossens}(1991)]{1991SoPh..134..247V}
	{van der Linden}, R.A.M.; {Goossens}, M.
	\newblock {The Thermal Continuum in Coronal Loops - Instability Criteria and
		the Influence of Perpendicular Thermal Conduction}.
	\newblock {\em \solphys} {\bf 1991}, {\em 134},~247--273.
	\newblock {\url{https://doi.org/10.1007/BF00152647}}.
	
	\bibitem[{Hood}(1992)]{1992PPCF...34..411H}
	{Hood}, A.W.
	\newblock {Instabilities in the solar corona}.
	\newblock {\em Plasma Physics and Controlled Fusion} {\bf 1992}, {\em
		34},~411--442.
	\newblock {\url{https://doi.org/10.1088/0741-3335/34/4/002}}.
	
	\bibitem[{Antolin} and {Froment}(2022)]{2022FrASS...920116A}
	{Antolin}, P.; {Froment}, C.
	\newblock {Multi-Scale Variability of Coronal Loops Set by Thermal
		Non-Equilibrium and Instability as a Probe for Coronal Heating}.
	\newblock {\em Frontiers in Astronomy and Space Sciences} {\bf 2022}, {\em
		9},~820116.
	\newblock {\url{https://doi.org/10.3389/fspas.2022.820116}}.
	
	\bibitem[{Tsiklauri} \em{et~al.}(2004){Tsiklauri}, {Nakariakov}, {Arber}, and
	{Aschwanden}]{2004A&A...422..351T}
	{Tsiklauri}, D.; {Nakariakov}, V.M.; {Arber}, T.D.; {Aschwanden}, M.J.
	\newblock {Flare-generated acoustic oscillations in solar and stellar coronal
		loops}.
	\newblock {\em \aap} {\bf 2004}, {\em 422},~351--355,
	\href{http://xxx.lanl.gov/abs/astro-ph/0402261}{{\normalfont
			[arXiv:astro-ph/astro-ph/0402261]}}.
	\newblock {\url{https://doi.org/10.1051/0004-6361:20040299}}.
	
	\bibitem[{Zavershinskii} \em{et~al.}(2021){Zavershinskii}, {Kolotkov},
	{Riashchikov}, and {Molevich}]{2021SoPh..296...96Z}
	{Zavershinskii}, D.; {Kolotkov}, D.; {Riashchikov}, D.; {Molevich}, N.
	\newblock {Mixed Properties of Slow Magnetoacoustic and Entropy Waves in a
		Plasma with Heating/Cooling Misbalance}.
	\newblock {\em \solphys} {\bf 2021}, {\em 296},~96,
	\href{http://xxx.lanl.gov/abs/2104.12652}{{\normalfont
			[arXiv:astro-ph.SR/2104.12652]}}.
	\newblock {\url{https://doi.org/10.1007/s11207-021-01841-1}}.
	
	\bibitem[{Zavershinskii} \em{et~al.}(2019){Zavershinskii}, {Kolotkov},
	{Nakariakov}, {Molevich}, and {Ryashchikov}]{2019PhPl...26h2113Z}
	{Zavershinskii}, D.I.; {Kolotkov}, D.Y.; {Nakariakov}, V.M.; {Molevich}, N.E.;
	{Ryashchikov}, D.S.
	\newblock {Formation of quasi-periodic slow magnetoacoustic wave trains by the
		heating/cooling misbalance}.
	\newblock {\em Physics of Plasmas} {\bf 2019}, {\em 26},~082113,
	\href{http://xxx.lanl.gov/abs/1907.08168}{{\normalfont
			[arXiv:astro-ph.SR/1907.08168]}}.
	\newblock {\url{https://doi.org/10.1063/1.5115224}}.
	
	\bibitem[{Zimovets} \em{et~al.}(2021){Zimovets}, {McLaughlin}, {Srivastava},
	{Kolotkov}, {Kuznetsov}, {Kupriyanova}, {Cho}, {Inglis}, {Reale}, {Pascoe},
	{Tian}, {Yuan}, {Li}, and {Zhang}]{2021SSRv..217...66Z}
	{Zimovets}, I.V.; {McLaughlin}, J.A.; {Srivastava}, A.K.; {Kolotkov}, D.Y.;
	{Kuznetsov}, A.A.; {Kupriyanova}, E.G.; {Cho}, I.H.; {Inglis}, A.R.; {Reale},
	F.; {Pascoe}, D.J.;  et~al.
	\newblock {Quasi-Periodic Pulsations in Solar and Stellar Flares: A Review of
		Underpinning Physical Mechanisms and Their Predicted Observational
		Signatures}.
	\newblock {\em \ssr} {\bf 2021}, {\em 217},~66.
	\newblock {\url{https://doi.org/10.1007/s11214-021-00840-9}}.
	
	\bibitem[{Kolotkov} and {Nakariakov}(2022)]{2022MNRAS.514L..51K}
	{Kolotkov}, D.Y.; {Nakariakov}, V.M.
	\newblock {A new look at the frequency-dependent damping of slow-mode waves in
		the solar corona}.
	\newblock {\em \mnras} {\bf 2022}, {\em 514},~L51--L55,
	\href{http://xxx.lanl.gov/abs/2205.05346}{{\normalfont
			[arXiv:astro-ph.SR/2205.05346]}}.
	\newblock {\url{https://doi.org/10.1093/mnrasl/slac054}}.
	
	\bibitem[{Prasad} \em{et~al.}(2021){Prasad}, {Srivastava}, and
	{Wang}]{2021SoPh..296..105P}
	{Prasad}, A.; {Srivastava}, A.K.; {Wang}, T.
	\newblock {Effect of Thermal Conductivity, Compressive Viscosity and Radiative
		Cooling on the Phase Shift of Propagating Slow Waves with and Without
		Heating-Cooling Imbalance}.
	\newblock {\em \solphys} {\bf 2021}, {\em 296},~105,
	\href{http://xxx.lanl.gov/abs/2104.07604}{{\normalfont
			[arXiv:astro-ph.SR/2104.07604]}}.
	\newblock {\url{https://doi.org/10.1007/s11207-021-01846-w}}.
	
	\bibitem[{Prasad} \em{et~al.}(2022){Prasad}, {Srivastava}, {Wang}, and
	{Sangal}]{2022SoPh..297....5P}
	{Prasad}, A.; {Srivastava}, A.K.; {Wang}, T.; {Sangal}, K.
	\newblock {Role of Non-ideal Dissipation with Heating-Cooling Misbalance on the
		Phase Shifts of Standing Slow Magnetohydrodynamic Waves}.
	\newblock {\em \solphys} {\bf 2022}, {\em 297},~5,
	\href{http://xxx.lanl.gov/abs/2112.04995}{{\normalfont
			[arXiv:astro-ph.SR/2112.04995]}}.
	\newblock {\url{https://doi.org/10.1007/s11207-021-01940-z}}.
	
	\bibitem[{Zavershinskii} \em{et~al.}(2020){Zavershinskii}, {Molevich},
	{Riashchikov}, and {Belov}]{2020PhRvE.101d3204Z}
	{Zavershinskii}, D.I.; {Molevich}, N.E.; {Riashchikov}, D.S.; {Belov}, S.A.
	\newblock {Nonlinear magnetoacoustic waves in plasma with isentropic thermal
		instability}.
	\newblock {\em \pre} {\bf 2020}, {\em 101},~043204.
	\newblock {\url{https://doi.org/10.1103/PhysRevE.101.043204}}.
	
	\bibitem[{Ledentsov}(2021)]{2021SoPh..296...74L}
	{Ledentsov}, L.
	\newblock {Thermal Trigger for Solar Flares I: Fragmentation of the Preflare
		Current Layer}.
	\newblock {\em \solphys} {\bf 2021}, {\em 296},~74,
	\href{http://xxx.lanl.gov/abs/2103.05664}{{\normalfont
			[arXiv:astro-ph.SR/2103.05664]}}.
	\newblock {\url{https://doi.org/10.1007/s11207-021-01817-1}}.
	
	\bibitem[{Nakariakov} \em{et~al.}(2017){Nakariakov}, {Afanasyev}, {Kumar}, and
	{Moon}]{2017ApJ...849...62N}
	{Nakariakov}, V.M.; {Afanasyev}, A.N.; {Kumar}, S.; {Moon}, Y.J.
	\newblock {Effect of Local Thermal Equilibrium Misbalance on Long-wavelength
		Slow Magnetoacoustic Waves}.
	\newblock {\em \apj} {\bf 2017}, {\em 849},~62.
	\newblock {\url{https://doi.org/10.3847/1538-4357/aa8ea3}}.
	
	\bibitem[{Kolotkov} \em{et~al.}(2019){Kolotkov}, {Nakariakov}, and
	{Zavershinskii}]{2019A&A...628A.133K}
	{Kolotkov}, D.Y.; {Nakariakov}, V.M.; {Zavershinskii}, D.I.
	\newblock {Damping of slow magnetoacoustic oscillations by the misbalance
		between heating and cooling processes in the solar corona}.
	\newblock {\em \aap} {\bf 2019}, {\em 628},~A133,
	\href{http://xxx.lanl.gov/abs/1907.07051}{{\normalfont
			[arXiv:astro-ph.SR/1907.07051]}}.
	\newblock {\url{https://doi.org/10.1051/0004-6361/201936072}}.
	
	\bibitem[{Kolotkov} \em{et~al.}(2021){Kolotkov}, {Zavershinskii}, and
	{Nakariakov}]{2021PPCF...63l4008K}
	{Kolotkov}, D.Y.; {Zavershinskii}, D.I.; {Nakariakov}, V.M.
	\newblock {The solar corona as an active medium for magnetoacoustic waves}.
	\newblock {\em Plasma Physics and Controlled Fusion} {\bf 2021}, {\em
		63},~124008,  \href{http://xxx.lanl.gov/abs/2111.02370}{{\normalfont
			[arXiv:astro-ph.SR/2111.02370]}}.
	\newblock {\url{https://doi.org/10.1088/1361-6587/ac36a5}}.
	
	\bibitem[{Kolotkov} \em{et~al.}(2020){Kolotkov}, {Duckenfield}, and
	{Nakariakov}]{2020A&A...644A..33K}
	{Kolotkov}, D.Y.; {Duckenfield}, T.J.; {Nakariakov}, V.M.
	\newblock {Seismological constraints on the solar coronal heating function}.
	\newblock {\em \aap} {\bf 2020}, {\em 644},~A33,
	\href{http://xxx.lanl.gov/abs/2010.03364}{{\normalfont
			[arXiv:astro-ph.SR/2010.03364]}}.
	\newblock {\url{https://doi.org/10.1051/0004-6361/202039095}}.
	
	\bibitem[{Zhugzhda}(1996)]{1996PhPl....3...10Z}
	{Zhugzhda}, Y.D.
	\newblock {Force-free thin flux tubes: Basic equations and stability}.
	\newblock {\em Physics of Plasmas} {\bf 1996}, {\em 3},~10--21.
	\newblock {\url{https://doi.org/10.1063/1.871836}}.
	
	\bibitem[{Duckenfield} \em{et~al.}(2021){Duckenfield}, {Kolotkov}, and
	{Nakariakov}]{2021A&A...646A.155D}
	{Duckenfield}, T.J.; {Kolotkov}, D.Y.; {Nakariakov}, V.M.
	\newblock {The effect of the magnetic field on the damping of slow waves in the
		solar corona}.
	\newblock {\em \aap} {\bf 2021}, {\em 646},~A155,
	\href{http://xxx.lanl.gov/abs/2011.10437}{{\normalfont
			[arXiv:astro-ph.SR/2011.10437]}}.
	\newblock {\url{https://doi.org/10.1051/0004-6361/202039791}}.
	
	\bibitem[{Belov} \em{et~al.}(2021){Belov}, {Molevich}, and
	{Zavershinskii}]{2021SoPh..296..122B}
	{Belov}, S.A.; {Molevich}, N.E.; {Zavershinskii}, D.I.
	\newblock {Dispersion of Slow Magnetoacoustic Waves in the Active Region Fan
		Loops Introduced by Thermal Misbalance}.
	\newblock {\em \solphys} {\bf 2021}, {\em 296},~122,
	\href{http://xxx.lanl.gov/abs/2107.10600}{{\normalfont
			[arXiv:astro-ph.SR/2107.10600]}}.
	\newblock {\url{https://doi.org/10.1007/s11207-021-01868-4}}.
	
	\bibitem[{Roberts}(2006)]{2006RSPTA.364..447R}
	{Roberts}, B.
	\newblock {Slow magnetohydrodynamic waves in the solar atmosphere}.
	\newblock {\em Philosophical Transactions of the Royal Society of London Series
		A} {\bf 2006}, {\em 364},~447--460.
	\newblock {\url{https://doi.org/10.1098/rsta.2005.1709}}.
	
	\bibitem[{Wang} \em{et~al.}(2007){Wang}, {Innes}, and
	{Qiu}]{2007ApJ...656..598W}
	{Wang}, T.; {Innes}, D.E.; {Qiu}, J.
	\newblock {Determination of the Coronal Magnetic Field from Hot-Loop
		Oscillations Observed by SUMER and SXT}.
	\newblock {\em \apj} {\bf 2007}, {\em 656},~598--609,
	\href{http://xxx.lanl.gov/abs/astro-ph/0612566}{{\normalfont
			[arXiv:astro-ph/astro-ph/0612566]}}.
	\newblock {\url{https://doi.org/10.1086/510424}}.
	
	\bibitem[{Kumar} and {Kumar}(2022)]{2022JApA...43...40K}
	{Kumar}, A.; {Kumar}, N.
	\newblock {Effect of heating-cooling imbalance on slow mode with time-dependent
		background temperature}.
	\newblock {\em Journal of Astrophysics and Astronomy} {\bf 2022}, {\em 43},~40.
	\newblock {\url{https://doi.org/10.1007/s12036-022-09824-9}}.
	
	\bibitem[{Del Zanna} \em{et~al.}(2021){Del Zanna}, {Dere}, {Young}, and
	{Landi}]{2021ApJ...909...38D}
	{Del Zanna}, G.; {Dere}, K.P.; {Young}, P.R.; {Landi}, E.
	\newblock {CHIANTI{\textemdash}An Atomic Database for Emission Lines. XVI.
		Version 10, Further Extensions}.
	\newblock {\em \apj} {\bf 2021}, {\em 909},~38,
	\href{http://xxx.lanl.gov/abs/2011.05211}{{\normalfont
			[arXiv:physics.atom-ph/2011.05211]}}.
	\newblock {\url{https://doi.org/10.3847/1538-4357/abd8ce}}.
	
	\bibitem[{Rosner} \em{et~al.}(1978){Rosner}, {Tucker}, and
	{Vaiana}]{1978ApJ...220..643R}
	{Rosner}, R.; {Tucker}, W.H.; {Vaiana}, G.S.
	\newblock {Dynamics of the quiescent solar corona.}
	\newblock {\em \apj} {\bf 1978}, {\em 220},~643--645.
	\newblock {\url{https://doi.org/10.1086/155949}}.
	
	\bibitem[{Dahlburg} and {Mariska}(1988)]{1988SoPh..117...51D}
	{Dahlburg}, R.B.; {Mariska}, J.T.
	\newblock {Influence of Heating Rate on the Condensational Instability}.
	\newblock {\em \solphys} {\bf 1988}, {\em 117},~51--56.
	\newblock {\url{https://doi.org/10.1007/BF00148571}}.
	
	\bibitem[{Ibanez S.} and {Escalona T.}(1993)]{1993ApJ...415..335I}
	{Ibanez S.}, M.H.; {Escalona T.}, O.B.
	\newblock {Propagation of Hydrodynamic Waves in Optically Thin Plasmas}.
	\newblock {\em \apj} {\bf 1993}, {\em 415},~335.
	\newblock {\url{https://doi.org/10.1086/173167}}.
	
	\bibitem[{Carbonell} \em{et~al.}(2006){Carbonell}, {Terradas}, {Oliver}, and
	{Ballester}]{2006A&A...460..573C}
	{Carbonell}, M.; {Terradas}, J.; {Oliver}, R.; {Ballester}, J.L.
	\newblock {Spatial damping of linear non-adiabatic magnetoacoustic waves in a
		prominence medium}.
	\newblock {\em \aap} {\bf 2006}, {\em 460},~573--581.
	\newblock {\url{https://doi.org/10.1051/0004-6361:20065528}}.
	
	\bibitem[{Iba{\~n}ez} and {Ballester}(2022)]{2022SoPh..297..144I}
	{Iba{\~n}ez}, M.H.; {Ballester}, J.L.
	\newblock {The Effect of Thermal Misbalance on Slow Magnetoacoustic Waves in a
		Partially Ionized Prominence-Like Plasma}.
	\newblock {\em \solphys} {\bf 2022}, {\em 297},~144.
	\newblock {\url{https://doi.org/10.1007/s11207-022-02071-9}}.
	
	\bibitem[{Testa} \em{et~al.}(2014){Testa}, {De Pontieu}, {Allred}, {Carlsson},
	{Reale}, {Daw}, {Hansteen}, {Martinez-Sykora}, {Liu}, {DeLuca}, {Golub},
	{McKillop}, {Reeves}, {Saar}, {Tian}, {Lemen}, {Title}, {Boerner},
	{Hurlburt}, {Tarbell}, {Wuelser}, {Kleint}, {Kankelborg}, and
	{Jaeggli}]{2014Sci...346B.315T}
	{Testa}, P.; {De Pontieu}, B.; {Allred}, J.; {Carlsson}, M.; {Reale}, F.;
	{Daw}, A.; {Hansteen}, V.; {Martinez-Sykora}, J.; {Liu}, W.; {DeLuca}, E.E.;
	et~al.
	\newblock {Evidence of nonthermal particles in coronal loops heated impulsively
		by nanoflares}.
	\newblock {\em Science} {\bf 2014}, {\em 346},~1255724,
	\href{http://xxx.lanl.gov/abs/1410.6130}{{\normalfont
			[arXiv:astro-ph.SR/1410.6130]}}.
	\newblock {\url{https://doi.org/10.1126/science.1255724}}.
	
	\bibitem[{Tajfirouze} \em{et~al.}(2016){Tajfirouze}, {Reale}, {Petralia}, and
	{Testa}]{2016ApJ...816...12T}
	{Tajfirouze}, E.; {Reale}, F.; {Petralia}, A.; {Testa}, P.
	\newblock {Time-resolved Emission from Bright Hot Pixels of an Active Region
		Observed in the EUV Band with SDO/AIA and Multi-stranded Loop Modeling}.
	\newblock {\em \apj} {\bf 2016}, {\em 816},~12,
	\href{http://xxx.lanl.gov/abs/1510.07524}{{\normalfont
			[arXiv:astro-ph.SR/1510.07524]}}.
	\newblock {\url{https://doi.org/10.3847/0004-637X/816/1/12}}.
	
	\bibitem[{Reale} \em{et~al.}(2019){Reale}, {Testa}, {Petralia}, and
	{Kolotkov}]{2019ApJ...884..131R}
	{Reale}, F.; {Testa}, P.; {Petralia}, A.; {Kolotkov}, D.Y.
	\newblock {Large-amplitude Quasiperiodic Pulsations as Evidence of Impulsive
		Heating in Hot Transient Loop Systems Detected in the EUV with SDO/AIA}.
	\newblock {\em \apj} {\bf 2019}, {\em 884},~131,
	\href{http://xxx.lanl.gov/abs/1909.02847}{{\normalfont
			[arXiv:astro-ph.SR/1909.02847]}}.
	\newblock {\url{https://doi.org/10.3847/1538-4357/ab4270}}.
	
	\bibitem[{Cranmer} and {Winebarger}(2019)]{2019ARA&A..57..157C}
	{Cranmer}, S.R.; {Winebarger}, A.R.
	\newblock {The Properties of the Solar Corona and Its Connection to the Solar
		Wind}.
	\newblock {\em \araa} {\bf 2019}, {\em 57},~157--187,
	\href{http://xxx.lanl.gov/abs/1811.00461}{{\normalfont
			[arXiv:astro-ph.SR/1811.00461]}}.
	\newblock {\url{https://doi.org/10.1146/annurev-astro-091918-104416}}.
	
	\bibitem[{Xia} and {Keppens}(2016)]{2016ApJ...823...22X}
	{Xia}, C.; {Keppens}, R.
	\newblock {Formation and Plasma Circulation of Solar Prominences}.
	\newblock {\em \apj} {\bf 2016}, {\em 823},~22,
	\href{http://xxx.lanl.gov/abs/1603.05397}{{\normalfont
			[arXiv:astro-ph.SR/1603.05397]}}.
	\newblock {\url{https://doi.org/10.3847/0004-637X/823/1/22}}.
	
	\bibitem[{Li} \em{et~al.}(2018){Li}, {Zhang}, {Peter}, {Chitta}, {Su}, {Xia},
	{Song}, and {Hou}]{2018ApJ...864L...4L}
	{Li}, L.; {Zhang}, J.; {Peter}, H.; {Chitta}, L.P.; {Su}, J.; {Xia}, C.;
	{Song}, H.; {Hou}, Y.
	\newblock {Coronal Condensations Caused by Magnetic Reconnection between Solar
		Coronal Loops}.
	\newblock {\em \apjl} {\bf 2018}, {\em 864},~L4,
	\href{http://xxx.lanl.gov/abs/1808.09626}{{\normalfont
			[arXiv:astro-ph.SR/1808.09626]}}.
	\newblock {\url{https://doi.org/10.3847/2041-8213/aad90a}}.
	
	\bibitem[{Mandrini} \em{et~al.}(2000){Mandrini}, {D{\'e}moulin}, and
	{Klimchuk}]{2000ApJ...530..999M}
	{Mandrini}, C.H.; {D{\'e}moulin}, P.; {Klimchuk}, J.A.
	\newblock {Magnetic Field and Plasma Scaling Laws: Their Implications for
		Coronal Heating Models}.
	\newblock {\em \apj} {\bf 2000}, {\em 530},~999--1015.
	\newblock {\url{https://doi.org/10.1086/308398}}.
	
	\bibitem[{Fleishman} \em{et~al.}(2021){Fleishman}, {Anfinogentov}, {Stupishin},
	{Kuznetsov}, and {Nita}]{2021ApJ...909...89F}
	{Fleishman}, G.D.; {Anfinogentov}, S.A.; {Stupishin}, A.G.; {Kuznetsov}, A.A.;
	{Nita}, G.M.
	\newblock {Coronal Heating Law Constrained by Microwave Gyroresonant Emission}.
	\newblock {\em \apj} {\bf 2021}, {\em 909},~89,
	\href{http://xxx.lanl.gov/abs/2101.03651}{{\normalfont
			[arXiv:astro-ph.SR/2101.03651]}}.
	\newblock {\url{https://doi.org/10.3847/1538-4357/abdab1}}.
	
\end{thebibliography}
\end{document}